%% file: main_seperate.tex
\documentclass[11pt, a4paper, logo, internal]{dm}
\usepackage{graphicx}%
\usepackage{multirow}%
\usepackage{amsmath,amssymb,amsfonts}%
\usepackage{amsthm}%
\usepackage{mathrsfs}%
\usepackage[title]{appendix}%
\usepackage{xcolor}%
\usepackage{textcomp}%
\usepackage{manyfoot}%
\usepackage{booktabs}%
\usepackage{algorithm}%
\usepackage{algorithmicx}%
\usepackage{algpseudocode}%
\usepackage{listings}%
\usepackage{hyperref}
\usepackage{lineno}

\theoremstyle{thmstyleone}%
\theoremstyle{thmstyletwo}%

\theoremstyle{thmstylethree}%
\input{defns}

\def\method{TeamPath}

\title{TeamPath: Building MultiModal Pathology Experts with Reasoning AI Copilots}

\begin{document}




\author[1,2,3,15]{Tianyu Liu}
\author[4,5,15]{Weihao Xuan}
\author[6,16]{Hao Wu}
\author[6,16]{Peter Humphrey}
\author[6,16]{Marcello DiStasio}
\author[6,16]{Mohamed Kahila}
\author[7,16]{Alfonso Garcia Tan}
\author[5]{Heli Qi}
\author[8]{Rui Yang}
\author[9]{Simeng Han}
\author[9]{Tinglin Huang}
\author[10]{Fang Wu}
\author[6]{Chen Liu}
\author[1,11]{Qingyu Chen}
\author[8,12,13]{Nan Liu}
\author[14]{Irene Li}
\author[1,11]{Hua Xu}
\author[1,2,*]{Hongyu Zhao}

\affil[1]{Interdepartmental Program in Computational Biology and Biomedical Informatics, Yale University}
\affil[2]{Department of Biostatistics, Yale University}
\affil[3]{Broad Institute of MIT and Harvard}
\affil[4]{Department of Complexity Science and Engineering, The University of Tokyo}
\affil[5]{Center for Advanced Intelligence Project, RIKEN}
\affil[6]{Department of Pathology, Yale University}
\affil[7]{Department of Anatomical Pathology, Singapore General Hospital}
\affil[8]{Center for Biomedical Data Science, Duke--NUS Medical School, Singapore, Singapore}
\affil[9]{Department of Computer Science, Yale University}
\affil[10]{Department of Computer Science, Stanford University}
\affil[11]{Department of Biomedical Informatics and Data Science, Yale University}
\affil[12]{Pre-hospital \& Emergency Research Center, Duke-NUS Medical School}
\affil[13]{Department of Biostatistics and Bioinformatics, Duke University}
\affil[14]{The Graduate School of Engineering, The University of Tokyo}
\affil[15]{These authors contributed equally to this work as leading authors.}
\affil[16]{These authors contributed equally to this work as human experts.}
\affil[*]{Corresponding author.}

\begin{abstract}
Advances in AI have introduced several strong models in computational pathology to usher it into the era of multi-modal diagnosis, analysis, and interpretation. However, the current pathology-specific visual language models still lack capacities in making the diagnosis with rigorous reasoning paths as well as handling divergent tasks, and thus, challenges of building AI Copilots for real scenarios still exist. Here we introduce TeamPath, an AI system powered by reinforcement learning and router-enhanced solutions based on large-scale histopathology multimodal datasets, to work as a virtual assistant for expert-level disease diagnosis, patch-level information summarization, and cross-modality generation to integrate transcriptomic information for clinical usage. We also collaborate with pathologists from Yale School of Medicine to demonstrate that TeamPath can assist them in working more efficiently by identifying and correcting expert conclusions and reasoning paths. \textcolor{red}{We also discuss the human evaluation results to support the reasoning quality from TeamPath.} Overall, TeamPath can flexibly choose the best settings according to the needs, and serve as an innovative and reliable system for information communication across different modalities and experts.
\end{abstract}


\maketitle

\section{Introduction}
\input{section_folder/introduction}
\section{Results}
\input{section_folder/results}

\section{Discussion}
\input{section_folder/discussion}

\section{Methods}
\input{section_folder/methods}

\bibliographystyle{unsrt}
\bibliography{sn-bibliography}

\newpage

\appendix
\input{section_folder/Appendix}

\end{document}

%% file: defns.tex
\usepackage{mathtools}
\usepackage[dvipsnames]{xcolor}
\usepackage[numbers, sort&compress, round]{natbib}
\usepackage{booktabs}
\usepackage{graphicx}
\usepackage{xfrac}
\usepackage{bbm}
\usepackage{changes}
\usepackage{rotating}
\usepackage{wrapfig}

\usepackage[most]{tcolorbox}
\usepackage{xparse}
\usepackage{adjustbox}
\usepackage{xspace}
\usepackage{changepage}
\usepackage{enumitem}
\usepackage{pifont}
\usepackage{ulem}
\usepackage{tocloft}
\usepackage[toc]{multitoc}
\usepackage{etoc}
\usepackage{dsfont}
\usepackage{dm-colors}
\usepackage{multirow}
\usepackage{minted}
\usepackage{caption}
\usepackage{soul}
\usepackage{float}
\usepackage{svg}
\usepackage{adjustbox}
\usepackage{setspace}
\svgsetup{inkscapelatex=false}
\usepackage{silence}  
\newcommand{\hide}[1]{}

\newcommand{\method}{SpatialAgent\,}

\pdftrailerid{redacted}

%% file: section_folder/introduction.tex
Pathological diagnosis is a complex yet essential component of clinical decision-making. Through the examination of whole-slide images (WSIs), physicians assess disease severity, evaluate the spatial distribution of malignant and healthy cells, and generate diagnostic reports or treatment recommendations \cite{song2023artificial, bera2019artificial, niazi2019digital, al2012digital}. However, this process is both time-intensive and labor-intensive, and its accuracy can be influenced by uncontrollable factors such as the physician’s workload, fatigue, and level of expertise \cite{zhang2024challenges}. Recent advances in Artificial Intelligence (AI) have demonstrated considerable promise in augmenting diagnostic workflows \cite{chen2024towards, lu2023towards, xu2024whole, ma2024generalizablepathologyfoundationmodel, Rannikko2025}. In particular, the deployment of foundation models for pathology not only reduces resource demands but also enables scalable, reproducible analysis. A deeper understanding of how these models generate diagnostic predictions, together with continued efforts to improve their mechanisms, is crucial for enhancing reliability and precision in clinical applications. This represents an emerging and important direction for future research.

Deconstructing the diagnostic workflow of pathologists provides critical insights into the role of AI in this domain. In practice, physicians analyze WSIs or regions of interest (ROIs) by examining selected patches, which might contain diagnostically relevant features. Localized assessments are then aggregated into a diagnostic report, which can support higher-level clinical investigations \cite{tran2025generating, liu2025spemo, shaikovski2025prism2} when paired with the corresponding ROIs. Meanwhile, visual–language models (VLMs) \cite{zhang2024vision} process paired image–text inputs, typically images with accompanying questions or instructions, and generate responses by integrating and aligning information across modalities. Inspired by this parallel, researchers have begun to develop VLMs specifically tailored for WSIs and pathology diagnosis. For instance, SlideChat \cite{Chen_2025_CVPR} and PathChat \cite{lu2024multimodal} were designed as copilots for pathology interpretation, while HistoGPT \cite{tran2025generating} can generate medical reports directly from histopathology images. Similarly, spEMO \cite{liu2025spemo} extends this capability by incorporating both molecular and pathological information for a stronger report generation capacity. Other pathology foundation models (PFMs), such as MUSK \cite{xiang2025vision} and PLIP \cite{huang2023visual}, leverage text–image alignment to improve embedding quality. Collectively, these domain-specific PFMs and VLMs have advanced applications in disease-state prediction, medical report generation and multimodal integration, establishing pathology-focused VLMs as a promising and rapidly progressing research direction.

Nevertheless, certain tasks in pathology diagnosis are inherently complex and require deliberate reasoning before actions can be taken. To address such challenges, foundation models must demonstrate robust reasoning capabilities. Conventional VLMs, however, often struggle with reasoning-oriented questions, even when trained on extensive datasets. Equipping VLMs with effective reasoning capacity thus remains a central challenge in the medical domain. At the same time, physicians play an indispensable role in the era of medical AI. They are not only domain experts but also important users and researchers for helping us correct errors made by AI models or interact with models to improve each other's performances. Consequently, developing effective strategies for human–AI collaboration, particularly in ways that enhance rather than replace physician expertise, is an urgent priority for advancing reliable and clinically meaningful solutions.

Fortunately, encouraging progress in reasoning has been achieved in other domains, such as mathematics and logic, through the training of large language models (LLMs, which are text-only) and VLMs. These advances are largely driven by reinforcement learning (RL) with/without chain-of-thought (CoT) supervision, which has demonstrated strong reasoning capabilities \cite{chu2025sft}. Importantly, such techniques can be adapted to medical applications, provided that domain-appropriate datasets are carefully constructed. Within pathology, several groups have begun to explore reasoning-oriented models for diagnostic tasks \cite{zhang2025patho, xu2025discovering, wu2025pathvlm}. However, current approaches tend to be technically homogeneous and insufficient to disentangle the contribution of the reasoning process from the final answer. Little attention has been given to analyzing errors produced by reasoning, which is critical for improving model reliability. Moreover, most existing models remain closed-source, which limit opportunities for rigorous evaluation in real-world clinical scenarios and hindering community-driven progress. To address these gaps, we aim to develop a high-precision reasoning model that not only generates accurate diagnostic predictions but also explicates its reasoning path. Such a model would serve as a trustworthy assistant to physicians, support more informed clinical decision-making, and ultimately contribute to alleviating patient burden while advancing the goals of precision medicine.

In this manuscript, we present \method{}, a framework that augments VLMs with multi-modal reasoning and a task-sensitive routing mechanism, enabling robust performance in several pathology-related tasks. Our approach begins with the careful selection of base models and the design of medical-specific prompts to curate high-quality, reasoning-enriched training data. Through comprehensive analyses, we demonstrate both the necessity of equipping VLMs with reasoning capabilities to address complex pathology tasks and the importance of constructing high-quality datasets for model success. We further showcase the effectiveness of \method{} across diverse downstream applications, including multi-modal pathology visual question answering (Pathology VQA) and caption summarization. By leveraging an LLM-driven router, \method{} dynamically selects the most suitable strategy to meet task requirements, functioning as a reliable and adaptive system. Importantly, we invite pathologists to evaluate the model’s reasoning pathways, thereby validating its practical utility as a medical assistant. Finally, we introduce a new task, known as spatial transcriptomic profiles generation, to assess the cross-modality generative ability of \method{}. Overall, \method{} provides a new avenue for integrative analyses that combine molecular and histopathological signatures.

%% file: section_folder/results.tex
\textbf{Dataset and Method Overview.} The curation of high-quality datasets is increasingly critical for advancing PFMs and VLMs, particularly in the era of multimodal reasoning and summarization. At the same time, careful attention must be paid to preventing data leakage to ensure unbiased evaluation of model performance. To this end, and leveraging prior data collection strategies, we distilled a subset of data from PathGen-1.6M \cite{sunpathgen}, which is a large-scale resource comprising nearly 10,000 WSIs and 1.6 million ROIs derived from TCGA data \cite{weinstein2013cancer}, for the usage in the finetuning stage with reinforcement learning. Reasoning data were constructed using COT templates generated based on the advanced reasoning model o4-mini \cite{oai4mini}, with subsequent quality validation performed by pathologists at Yale School of Medicine. Importantly, this dataset does not overlap with the benchmark testing set used for Pathology VQA evaluation \cite{he2020pathvqa}, namely PathMMU \cite{sun2024pathmmu}, which contains ROIs paired with questions across five diagnostic categories and represents one of the most advanced evaluation sources. In addition, another subset distilled from PathGen-1.6M was curated as the testing dataset for the ROI summarization task. To assess performance in cross-modality generation, we leveraged HEST-1K \cite{jaume2024hest} and STImage1K4M \cite{chen2024stimage}, two multi-omic histopathology collections to assess the prediction of transcriptomic profiles as molecular signatures from ROIs. These two datasets are used to construct training, validation, and test sets. The overall data preprocessing workflow and sample sizes are summarized in Extended Data Figure \ref{supfig:dataflowchart}.

\begin{figure}
    \centering
    \includegraphics[width=1\linewidth]{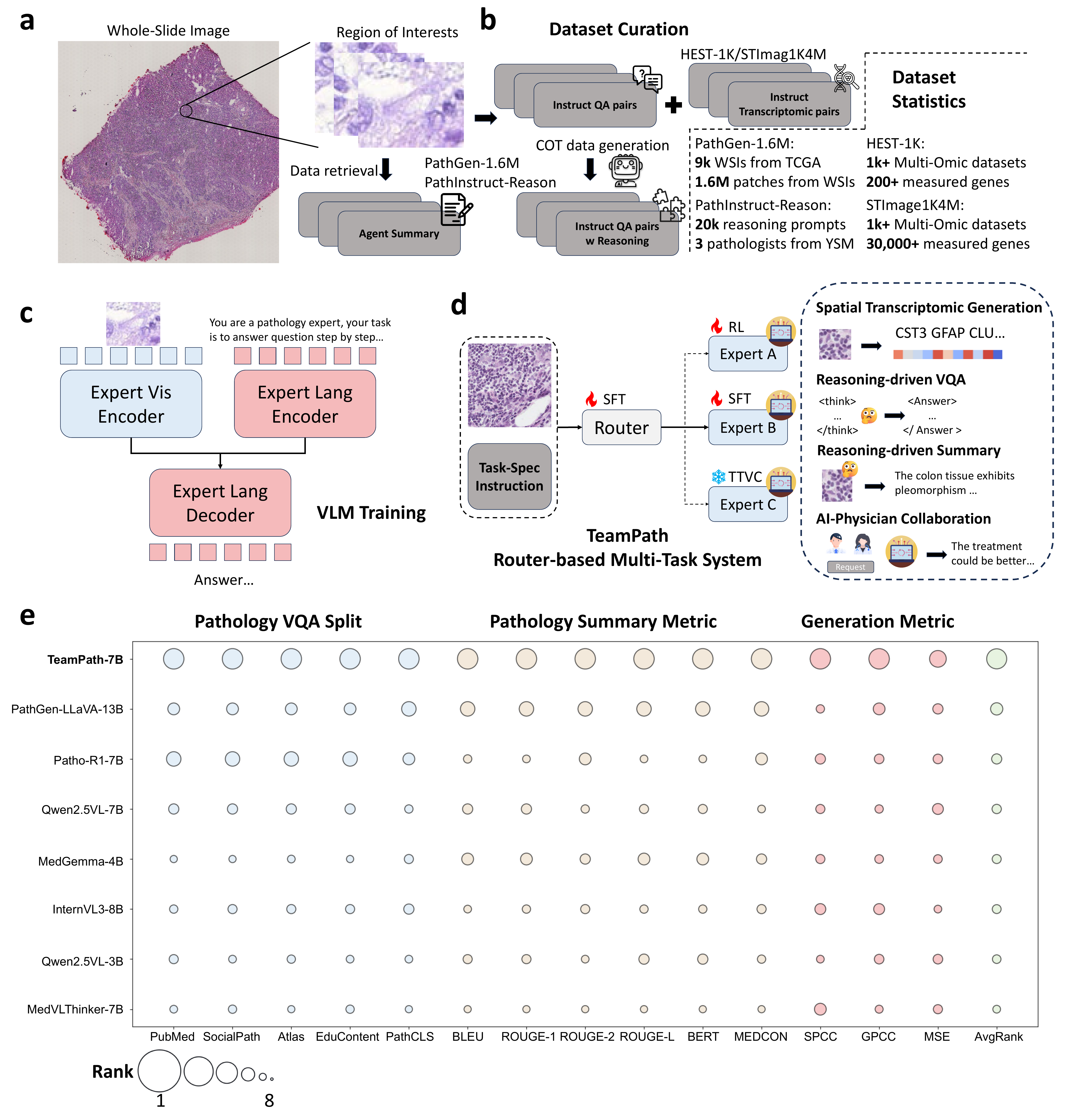}
    \caption{Landscape of \method{} (a) Steps of dataset curation. We extract image-text pairs from a processed TCGA dataset (PathGen-1.6M). (b) Word cloud visualization of ROI captions (upper) and questions (bottom). (c) The core visual language model architecture of \method{}. (d) \method{} as a system with an LLM-enhanced router (with 100\% accuracy in choosing the correct approach) and the corresponding capacities in various downstream applications. The logo fire means that we need to adjust the parameters of models, and the logo snowflake means that we do not change the parameters. (e) Overall ranking list of different methods across tasks and metrics. \textcolor{red}{A lower rank (also a larger bubble) means a better method.}}
    \label{fig:overall figure}
\end{figure}

The overall process of dataset curation and model training is summarized in Figures \ref{fig:overall figure} (a)-(d). \method{} emerges as a robust multimodal AI assistant for both disease analysis and modality generation. To refine its reasoning capabilities, we employ Group Relative Policy Optimization (GRPO) \cite{guo2025deepseek} to finetune the base model (the default setting is Patho-R1-7B), thereby enhancing its ability to perform reasoning over pathology images. With this capacity for structured reasoning, \method{} demonstrates strong performance in addressing Pathology VQA tasks, as shown in our comprehensive benchmarking analysis. Importantly, the model also maintains high performance on tasks where reasoning is less critical, such as image summarization (known as caption generation) and cross-modality generation. This adaptability enables \method{} to support task-specific optimization through either reinforcement learning or supervised finetuning. In collaboration with expert pathologists, we further demonstrate that \method{} can function as a clinical copilot, assisting in applications such as correcting inaccurate conclusions and identifying flawed reasoning paths. Taken together, \method{} advances both biomedical research and clinical practice in histopathology analysis. Finally, a comparative summary of task- and metric-specific rankings, shown in Figure \ref{fig:overall figure} (e), demonstrates the superior performance of \method{} across multiple dimensions.

\textbf{\method{} improves the performance of ROI-level assessment with reasoning ability.} The increasing complexity of histopathology image analysis presents significant challenges for developing expert-level VLMs. One particularly demanding setting is Pathology VQA, which requires models to correctly respond to questions grounded in histopathology images. Unlike traditional classification tasks (e.g., disease-state classification or cancer cell identification), Pathology VQA involves a broader and more complex range of scenarios \cite{he2020pathological} and demands higher accuracy in answer production. To evaluate model performances under this setting, we employ the recently published PathMMU dataset, which includes VQA pairs spanning five categories, ranging from expert-annotated questions to images from social media. Importantly, PathMMU is excluded from the training data of all evaluated models to ensure fairness. Reflecting the real-world requirements faced by pathologists, we emphasize the need for high-quality, fine-grained answers that integrate multimodal information and contribute meaningfully at the clinical level. Our baseline comparisons encompass (1) general-domain VLMs, including o4-mini, GPT-4o \cite{hurst2024gpt}, Qwen2.5VL-3B, Qwen2.5VL-7B \cite{team2024qwen2}, and InternVL3-8B \cite{zhu2025internvl3}; (2) medical-domain VLMs, including MedGemma-4B \cite{sellergren2025medgemma} and MedVLThinker-7B \cite{huang2025medvlthinker}; (3) pathology-specific VLMs, including PathGen-LLaVA-13B \cite{sunpathgen} and Patho-R1-7B \cite{zhang2025patho}; and (4) a random-answer baseline. Model performance is assessed by computing accuracy relative to expert-generated answers within PathMMU, enabling a rigorous and fair benchmarking analysis. \textcolor{red}{The comparison of sample size used for training and testing is shown in Extended Data Table \ref{suptab: samplesize}, and we can see that the number of images used in testing is large enough to support a general conclusion.}

\begin{figure}
    \centering
    \includegraphics[width=1\linewidth]{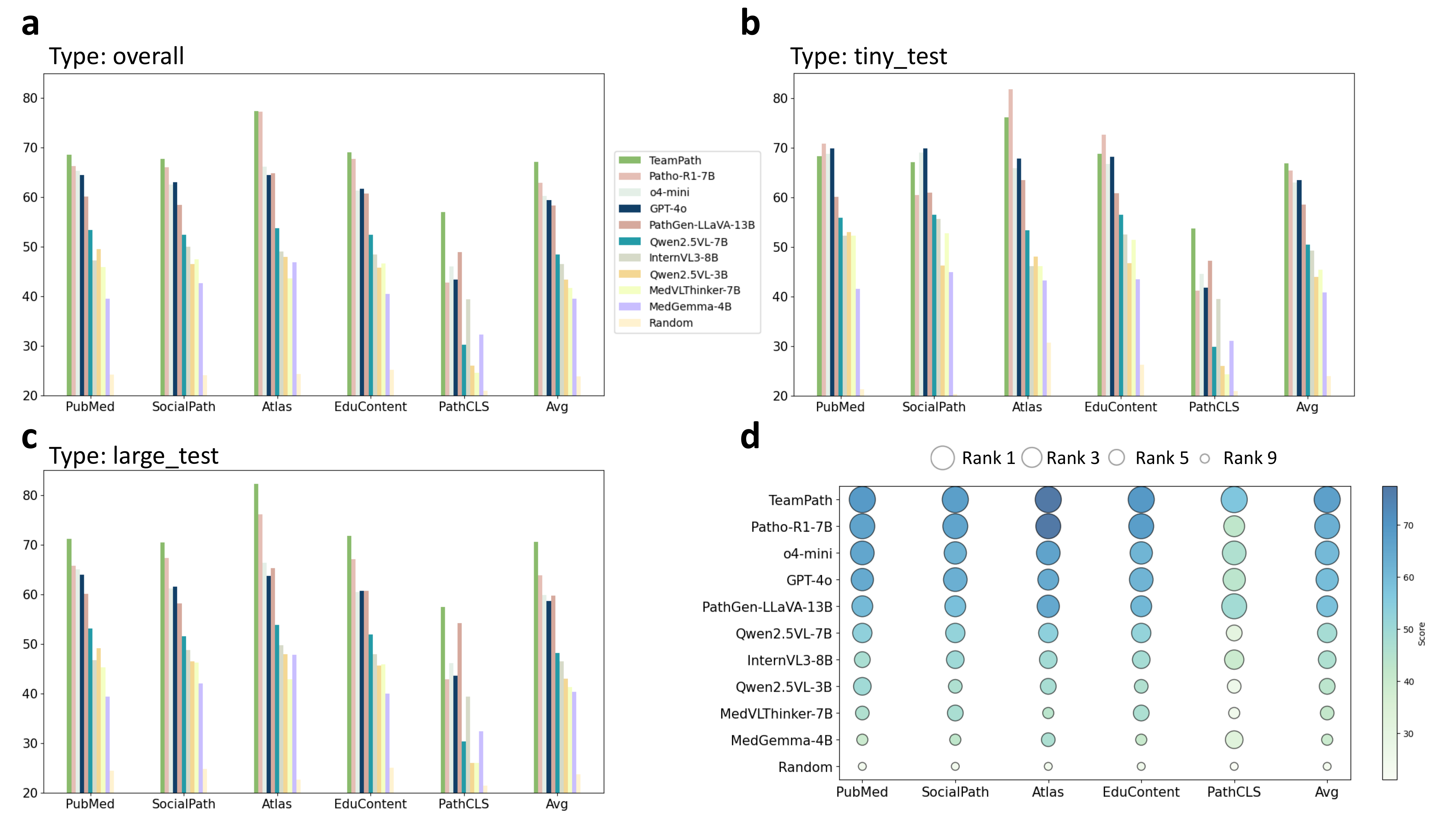}
    \caption{Benchmarking results with PathMMU for the pathology VQA task. We note that since we did not have information about the testing setting of PathGen-LLaVA-13B, we used results reported by the model creators in \cite{sunpathgen}. (a) Accuracy across different categories of all selected methods with all samples. (b) Accuracy across different categories of all selected methods with samples from a tiny set. (c) Accuracy across different categories of all selected methods with samples from a large set. (d) Joint visualization with accuracy and ranking information for all selected methods. The darker the bubble color, the higher the model score; The larger the bubble shape, the lower the model ranking.}
    \label{fig:pathmmu bench}
\end{figure}

Figures \ref{fig:pathmmu bench} (a)-(c) show our benchmarking results across different categories, including PubMed, SocialPath, Atlas, EduContent, and PathCLS. PathMMU also pre-defines different sample types, and ``overall" represents all testing samples in the selected category, ``tiny\_test" represents testing samples used for expert evaluation, and ``test'' represents the rest of the samples. We find that \method{} outperforms all other baseline models, including domain-expert VLMs with similar or larger parameter size, such as Patho-R1-7B and PathGen-LLaVA-13B, in nearly all evaluations. \textcolor{red}{\method{} also performs better than strong general VLMs, such as o4-mini and GPT-4o, further demonstrating the strength of expert models in addressing medical challenges. Moreover, o4-mini and GPT-4o still perform better than most of the selected baselines, indicating they possess a certain level of understanding of pathological knowledge.} Other general VLMs and medical VLMs performed poorly in this task. We further visualize the comprehensive benchmarking analysis, including ranking and accuracy of each method with all samples in Figure \ref{fig:pathmmu bench} (d), which shows that \method{} also has the lowest rank by considering all categories jointly. Therefore, our experiment results show that introducing reasoning capacities to build pathology-expert VLMs can enhance their ability in making diagnoses, and thus \method{} can serve as a strong performer for the key feature identification and content understanding of ROIs.

\textcolor{red}{To examine the performance of \method{} specific to disease diagnosis, we select diagnosis-related questions in PathMMU as a subset to make a comparison. According to Extended Data Figures \ref{supfig:diaginfo} (a)-(b), \method{} performs better than the second-best baseline Patho-R1-7B as well as random guessing in handling disease-diagnosis-related queries, and the improvement is consistent across most of the disease categories in both the tiny group and the large group. Therefore, \method{} can also improve diagnostic accuracy after training.}

To obtain a more intuitive understanding of the key contributions of \method{} following reinforcement learning training, we selected two case studies where \method{} provided the correct answer while other models failed to make accurate judgments. 

\begin{figure}
    \centering
    \includegraphics[width=1\linewidth]{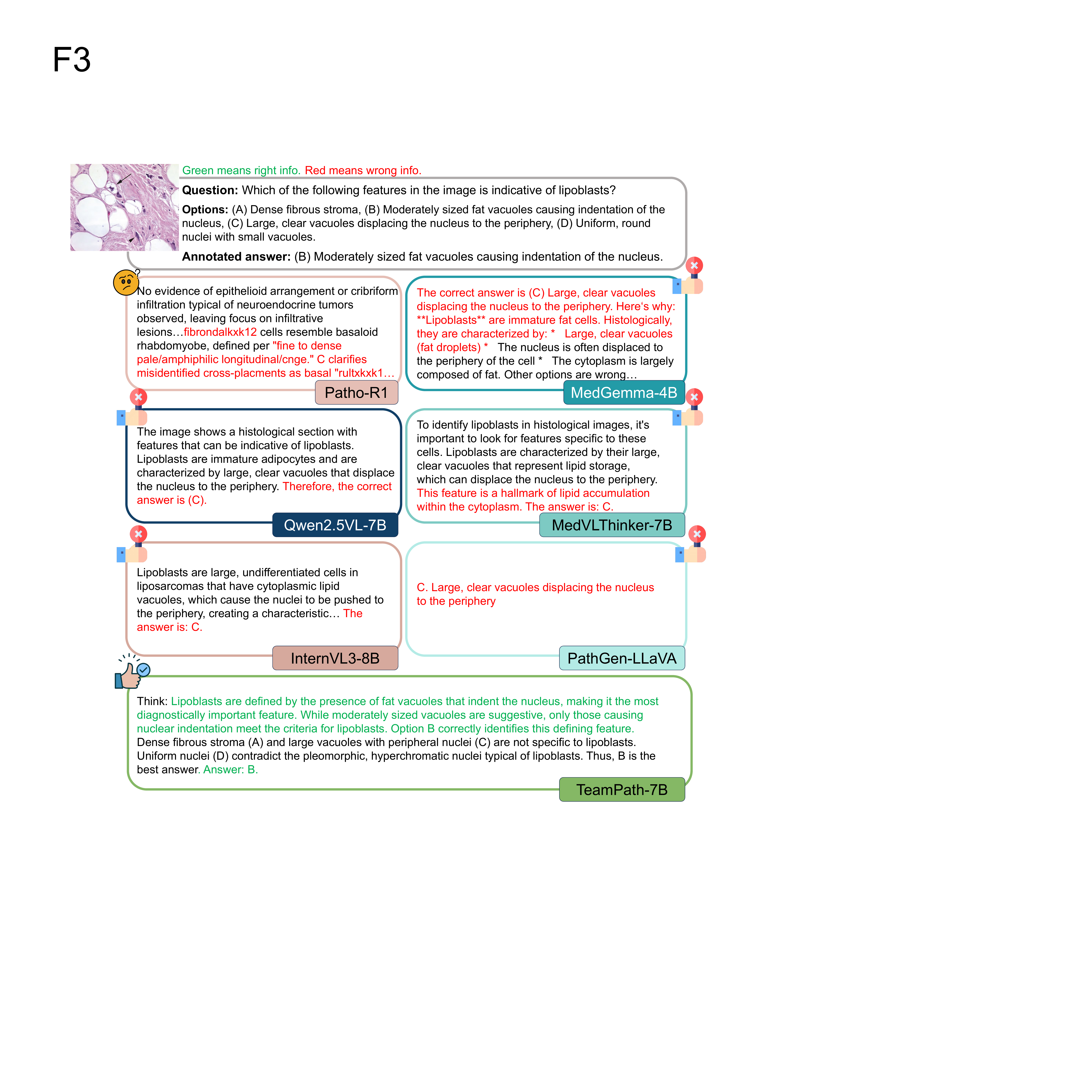}
    \caption{Case study (topic: synaptophysin, which is a precursor cell that develops into an adipocyte (fat cell)) based on the outputs from different models. We highlight the correct information with green text and incorrect information with red text. For the models with errors, we consider two cases. The first case is a wrong answer, and the second case is a confused reasoning path.}
    \label{fig:casestudy1}
\end{figure}

Figure \ref{fig:casestudy1} highlights the importance of precise morphological criteria in recognizing lipoblasts. While several models incorrectly selected option C, describing large, clear vacuoles displacing the nucleus to the periphery, a hallmark of mature adipocytes. We found that \method{} correctly identified option B as the defining feature of lipoblasts \cite{hisaoka2014lipoblast}. Lipoblasts are diagnostically recognized by the presence of moderately sized cytoplasmic fat vacuoles that indent or scallop the nucleus, a distinction that separates them from both mature adipocytes and other stromal features. By emphasizing nuclear indentation rather than displacement, \method{} demonstrated accurate pathological reasoning aligned with standard diagnostic criteria. This correctness not only underscores the reliability of \method{} in differentiating subtle histologic features but also highlights the critical nuance needed in distinguishing malignant lipoblastic cells from benign adipocytic processes. Moreover, Extended Data Figure \ref{supfig:casestudy2} demonstrates that \method{} correctly identified synaptophysin as the targeted marker in the immunohistochemical stain of section A. The brown, cytoplasmic staining pattern observed is a hallmark of synaptophysin, which is widely used as a marker of neuroendocrine differentiation. While other models misclassified the stain as estrogen receptor or S100 protein, \method{} distinguished the subtle morphological and staining features that separate synaptophysin from nuclear markers like estrogen receptor or more diffuse proteins such as S100. This highlights both the accuracy and interpretive strength of \method{} in immunohistochemistry tasks, particularly in recognizing marker-specific staining patterns and avoiding common pitfalls that lead to misclassification. We also note that previous pathology expert models have obvious shortcomings, such as Patho-R1-7B's garbled output and PathGen-LLaVA's lack of interpretable diagnostic outputs. Instead, \method{} can make correct identification supported by comprehensive explanations, explained in the information provided by the reasoning paths.

We also explored the contributions of different training strategies and highlighted the importance of selecting base models based on a set of ablation studies, discussed in Appendix \ref{append: sftrl} and Extended Data Figures \ref{supfig:sftrl compare} (a)-(d), as well as in Appendix \ref{append: wordandchoice} and Extended Data Figure \ref{supfig:openclose compare} for the data ablation study.

\begin{figure}
    \centering
    \includegraphics[width=1\linewidth]{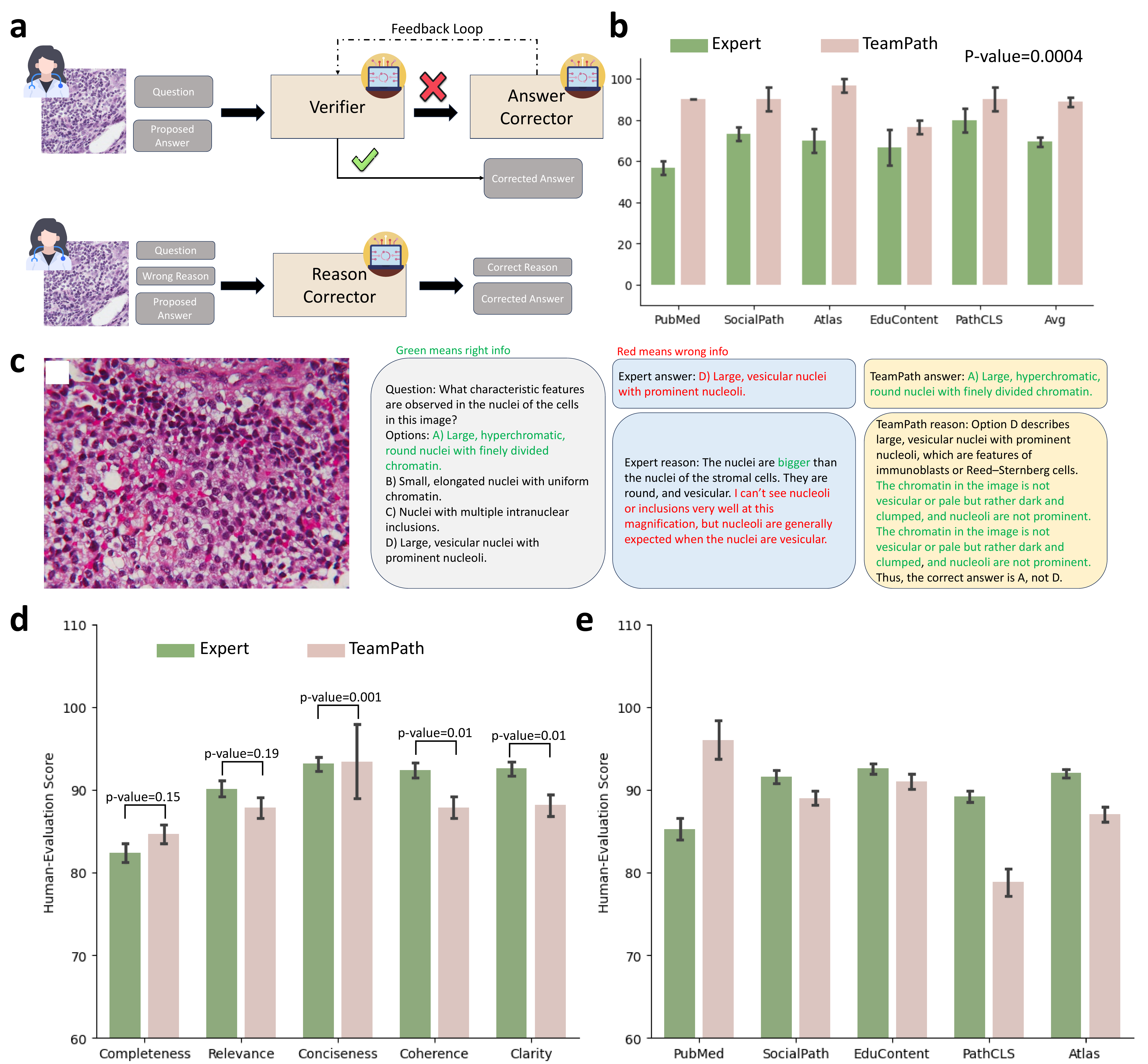}
    \caption{Results of using \method{} as the answer corrector/reason corrector. \method{} can work with pathologists together to improve the decision accuracy and provide explainable reasons to support the decision. (a) The illustration of self-verification/correction steps for both answers and reasoning paths. (b) Accuracy before and after correction based on selected samples from PathMMU. We report the average scores and standard deviation across three experts. The test is a one-sided Wilcoxon Rank-sum test. (c) A case study to demonstrate the power of \method{} as an AI assistant. \textcolor{red}{(d) Human evaluation results from three pathologist to compare the reasoning paths from humans and \method{} by metrics, and p-values based on the Wilcoxon Rank-sum test are attached. (e) Human evaluation results by problem categories.}}
    \label{fig:corrector performance}
\end{figure}

\textbf{\method{} acts as a Copilot in the pathologists-AI collaboration system.} Beyond demonstrating the capacity of \method{} in handling VQA sets as a pathology expert, we further explore its potential as an AI-assisted collaborator \cite{mialon2023gaia, liu2025towards}. An effective copilot should not only provide accurate responses to user queries but also reduce the effort required to resolve them, thereby saving both time and cost. To this end, we designed an algorithm in \method{} with test-time verification and correction (TTVC) \cite{liu2025towards} and engaged expert pathologists from Yale School of Medicine (YSM) to collaborate with \method{} in analyzing histopathology images and generating answers on demand. \textcolor{red}{Our TTVC pipeline operates at inference time and incurs additional computation through iterative verification and correction loops to improve answer quality, which aligns with the core principle of test-time scaling (TTS) \cite{snell2024scaling}: trading more computational resources at test time for better results, but we also extend TTS to the collaboration of different models. TeamPath-7B serves for correction and o4-mini serves for verification.} Specifically, we randomly subsampled 10 question–image pairs from each category within the PathMMU ``tiny\_test'' set and examined two capacities: (1) the ability of \method{} to act as an auto-verifier or auto-corrector for incorrect expert assessments, and (2) the ability of \method{} to revise and correct reasoning pathways when human experts fail to provide accurate answers. When performing the TTVC, we prepared the inputs as the images, questions, and reasoning from experts, and performed verification and correction based on \method{}. \textcolor{red}{Given the pathologists’ availability and the nature of the research, we have adopted an offline collaboration approach. We collect the pathologists’ answers and reasoning processes and use them as input to complete the two tasks described above using \method{}.} The overall paradigm for these two tasks is summarized in Figure \ref{fig:corrector performance} (a). Through this study, we aim to establish future paradigms of human–AI collaboration in biomedical research and clinical practice, highlighting the role of \method{} as a reliable and strong copilot. 

We jointly compared the expert-provided results with those corrected by \method{} and visualized the corresponding accuracies in Figure \ref{fig:corrector performance} (b). Our analysis shows that \method{} significantly improves accuracy across all PathMMU categories (p-value = 0.004), demonstrating that its corrective contribution is consistent and robust regardless of the source of pathology ROIs or questions. Notably, even in categories where expert performance is relatively low, such as PubMed, \method{} achieves substantial gains. These improvements demonstrate the effectiveness of \method{} as a corrector, as reflected by the observed accuracy differences. To further illustrate this capability, we conducted a case study (Figure \ref{fig:corrector performance} (c)) in which the expert provided an incorrect answer, whereas \method{} generated the correct response with an improved reasoning path. In this example, the task involved identifying characteristic nuclear features within the image. The expert’s reasoning correctly accounted for cell size but overlooked nucleolar details, leading to an erroneous conclusion. In contrast, \method{} integrated multiple features, including nuclear size, shape, staining depth, and prior knowledge of the cancer cell line, to eliminate incorrect options and arrive at the correct decision. Moreover, \method{} was also able to revise flawed reasoning paths when experts could not provide an answer (e.g., ``I do not know"), as shown in Extended Data Figure \ref{supfig:correction example}. In summary, through collaborative evaluation with pathologists, we demonstrate the capacity of \method{} to not only fix erroneous answers but also provide explicit reasoning steps, thereby enhancing both the transparency and interpretability of model-assisted pathology analysis.

\textcolor{red}{To directly compare the quality of reasoning paths from human experts and \method{}, we invited three pathologists with different backgrounds to evaluate these paths from five different metrics, including Completeness, Relevance, Conciseness, Coherence, and Clarity. Details of our grading criteria can be found in the Methods section. All scores range from 0 to 100, and a higher score represents a better quality reasoning path. We analyzed the reasoning paths generated by a top-scoring pathologist and \method{}. According to Figure \ref{fig:corrector performance} (d), the quality of reasoning paths from two sources is similar in terms of quality, and paths from \method{} have a higher Completeness score and a slightly higher Conciseness score, while paths from the human experts have higher Relevance, Coherence, and Clarify scores. After performing statistical testing analysis with the Wilcoxon Rank-sum test, the difference of most metrics is also not significant, marked in the image (Only the Conciseness score has a p-value 0.001, and other scores all have p-values$>$0.005). Therefore, we believe that the quality of the produced reasoning paths is comparable. Moreover, based on our results shown in Figure \ref{fig:corrector performance} (e), we find that methods have different preferences in problem categories. Here \method{} performs better in the questions from the PubMed category, while the strength of the human expert is more obvious in the questions from the PathCLS category. One possible reason is that most of the problems in PubMed come from scientific papers, which are quite similar to the model’s training environment. By correcting the pathologist’s flawed reasoning and incorporating different pathologists into a double-blind experiment to make a judgment, our study reveals the quality of reasoning paths generated by \method{} and explains its unique advantages in specific problems.}

\textcolor{red}{We also considered comparisons in terms of efficiency. We recorded the time it took human pathologists to complete the two tasks—answering questions and scoring reasoning steps. As shown in Extended Data Figure \ref{supfig:time compare}, \method{} can complete the answering of questions and the correction of reasoning steps within a single day, whereas even the fastest human pathologists require at least four days (time reported by humans) to reach a definitive conclusion. Therefore, the AI-based method also makes a significant contribution to improving the efficiency of medical problem-solving.}

\textcolor{red}{Here we provided more detailed analyses for understanding the case that \method{}'s discrepancies arising in correcting the feedback from different pathologists. We first compared the Pearson Correlation Coefficients (PCCs) among the three pathologists' accuracy. Extended Data Figure \ref{supfig:human result show} (a) shows that the decision-making of our two pathologists is relatively consistent, while their accuracy rates on specific datasets also differ. Therefore, we can ensure a certain degree of diversity in expert selection used for the human-AI collaboration experiment. Moreover, we also compared the number of corrected samples made by \method{} for different pathologists, and Extended Data Figure \ref{supfig:human result show} (b) shows that the number of corrections made by different experts is relatively similar across most datasets, with the sole exception being the SocialPath dataset. Since this dataset is primarily sourced from social media, the complex data pattern can pose challenges to our designed system. Our experiments demonstrate that \method{} contributes to improving the performance across different pathologists, thereby exhibiting relatively broad applicability. Furthermore, we also report the PCCs of pathologists' scores for reasoning paths. According to Extended Data Figure \ref{supfig:humaneval corr}, we also find obvious divergence for the decisions from different pathologists, and three pathologists show a positive PCC only in their Clarity scores. All of the demographic and training information for pathologists can be found in Supplementary Table 1. The pathologists we recruit all have at least three years of professional experience and have been trained and worked at leading medical institutions around the world. As a result, the quality of our pathologists is high. At the same time, we have observed that different pathologists specialize in different diseases, which helps explain the variations in their judgment to some extent.}

\textcolor{red}{We have also performed ablation studies for the verifier with three different choices (using the corrector, o3 \cite{oai4mini}, and o4-mini). Extended Data Figure \ref{supfig:verifier abla} shows that using o4-mini can achieve the best performance on average, while it can also reduce the cost compared with using o3 or a more advanced model. Since determining whether an answer is correct and correcting incorrect answers are two distinct types of problems, it makes sense to use different models for each to achieve the optimal performance. Therefore, o4-mini is selected here to perform verification.}

\textbf{\method{} performs better in summarizing the key information from histopathology images.} In practical applications, pathology image analysis often extends beyond generating correct answers and reasoning steps to encompass the extraction of important image features for macroscopic or high-level descriptions. To evaluate this capability, we designed experiments aimed at summarizing histopathology information from different ROIs, thereby assessing the capacity of \method{} to capture and convey high-level image content. \textcolor{red}{For this purpose, we constructed a testing dataset by subsampling 3,000 images and their corresponding captions from PathGen-1.6M. In their setting, the original image caption is enhanced by multi-agent (GPT-4V as backbone) collaboration. These captions were further annotated to include tissue- and disease-state information based on prompting Deepseek-R1 \cite{guo2025deepseek} to extract the answer from the input caption (extraction accuracy in the testing dataset is 70\% for disease-state labels and 94\% for tissue labels, validated by human experts and Deepseek-V3.2 \cite{liu2025deepseek}, and the Deepseek-R1's extraction error stems from classifying samples with mild inflammation as healthy samples).} To support training, we curated a separate dataset of 50,000 images, ensuring no overlap with the testing set. For benchmarking, we employed the same set of baseline models used in the Pathology VQA experiments. Model performance was evaluated using multiple similarity metrics between generated summaries and reference captions, including BLEU \cite{papineni2002bleu}, ROUGE-1/2/L \cite{lin2004rouge}, BERTScore \cite{zhangbertscore}, and MEDCON \cite{soldaini2016quickumls}. All metrics were in 0–100 range, with higher values indicating better performance.

\begin{figure}[H]
    \centering
    \includegraphics[width=0.9\linewidth]{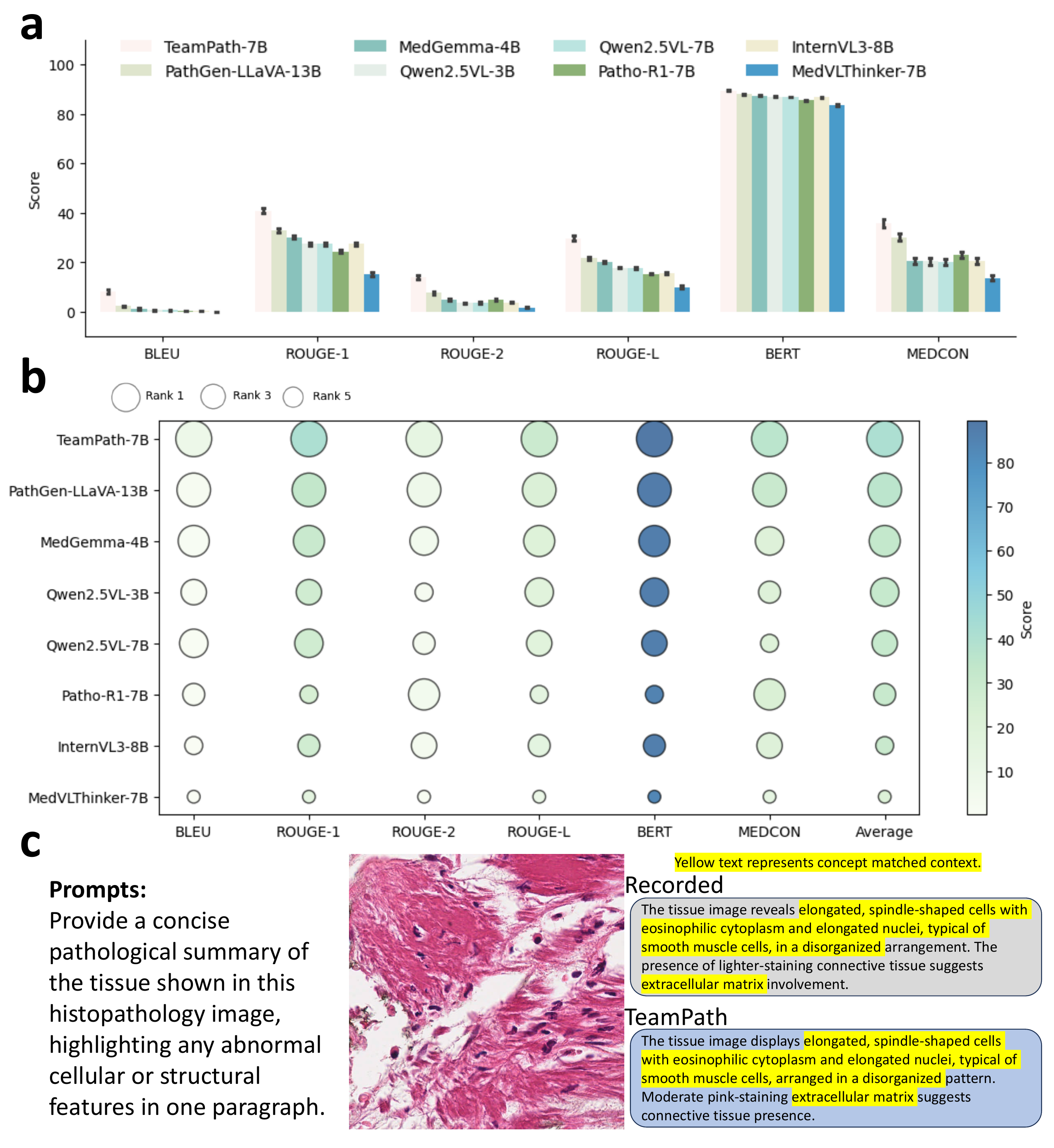}
    \caption{Benchmarking results of the caption summary task. (a) Performances of different methods for summarizing the caption based on ROI-level information across all metrics. We report the average scores and scaled standard deviation (0.1*sd) with all samples in the testing set. (b) Joint visualization with metric scores and ranking information for all selected methods. The darker the bubble color, the higher the model score; \textcolor{red}{The larger the bubble shape, the lower the model ranking.} (c) A case study of caption summary generation based on \method{}.}
    \label{fig:summary bench}
\end{figure}

Figure \ref{fig:summary bench} (a) compares the performance of \method{} with other VLMs across all selected metrics on the testing set. \method{} consistently outperforms the baselines across every metric, demonstrating its strength in generating summaries that align closely with reference annotations in both content and structure. To provide a holistic assessment, we further visualized the aggregated rankings and average scores of all methods in Figure \ref{fig:summary bench} (b), which highlights the leading performance of \method{} across the joint set of evaluation metrics. Recognizing that performance may vary by sample source, we also examined model performance across specific tissue and disease contexts. \textcolor{red}{We first analyze the produced categories in the testing dataset and illustrate the distribution of top 10 categories in Extended Data Figure \ref{supfig:dis and tissue category}. Among these labels, we select the top 1 label in each class, including adenocarcinoma and breast tissue, for further investigation.} Extended Data Figure \ref{supfig:summary by category} (a) reports ROUGE-L and BERT scores for samples from patients with adenocarcinoma, while Extended Data Figure \ref{supfig:summary by category} (b) shows results for breast tissue samples. In both cases, \method{} maintains superior performance compared with competing baselines. As an illustrative case study, Figure \ref{fig:summary bench} (c) presents an example output from \method{}, which accurately captures key organizational and pathological features—such as elongated spindle-shaped cells with eosinophilic cytoplasm and elongated nuclei, characteristic of smooth muscle cells. By contrast, outputs from baseline models (Extended Data Figure \ref{supfig:summary info}) contain less precise descriptions and, in some cases, incorrect content, further underscoring the advantages of \method{} in summarization tasks.

Therefore, we conclude that \method{} demonstrates as a strong performer in providing the high-level interpretations with pathology features of assigned ROIs.

\textbf{\method{} introduces new modalities with a cross-modality generation pipeline.} Building on our previous research and the existing literature, we observe that current histopathology image analyses primarily rely on textual and visual interpretations. However, given the breadth of biological signatures that can contribute to disease modeling and diagnosis, there is a clear opportunity to design new pipelines that integrate molecular information with histopathology features. Such integration can enable the generation of new modalities and provide deeper insights into cellular heterogeneity, lineage tracing, and disease mechanisms \cite{chen2025visual, song2024analysis}. Therefore, we finetune \method{} using paired histopathology images and transcriptomic profiles generated with the Visium technology \cite{visiumtech}, a platform for spatial transcriptomics (ST). Each ST spot includes a histopathology image as background and a corresponding gene expression profile. Inspired by Cell2Sentence \cite{levine2024cell2sentence} and Loki \cite{chen2025visual}, we convert gene expression profiles into ranked gene lists, ordering genes from highest to lowest expression. The task is then to generate these “spot sentences” and map them back into the transcriptomic space. For training and evaluation, we use two of the largest public datasets: HEST-1K (invasive ductal carcinoma, IDC) and STImage1K4M (brain tissue). HEST-1K includes a broad range of cancer datasets, whereas STimage1K4M contains samples from both disease and normal tissues, thereby enhancing the modeling of ST data. Baseline models for this task include the same VLMs evaluated in the Pathology VQA setting, supplemented with Cell2Sentence-1B. Model performance is assessed using Spot-level Pearson Correlation Coefficient (SPCC), Gene-level Pearson Correlation Coefficient (GPCC), and mean squared error (MSE). For SPCC and GPCC, higher values indicate better performance, whereas lower MSE values reflect higher accuracy.

\begin{figure}
    \centering
    \includegraphics[width=1\linewidth]{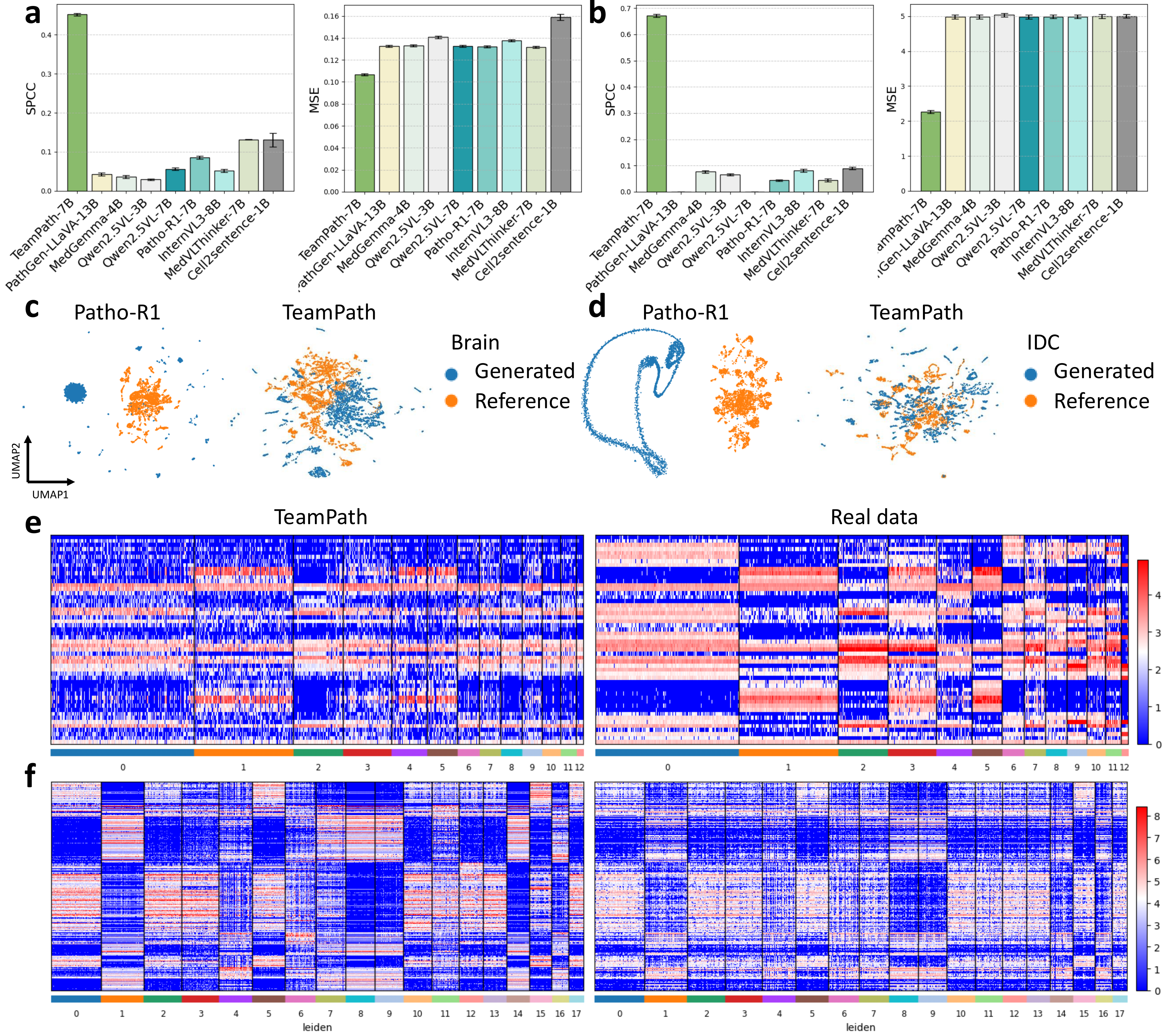}
    \caption{Evaluation of model performances for transcriptomic profile generation. (a) SPCC (higher is better) and MSE (lower is better) scores across different methods for the brain tissue. We report the average scores and scaled standard deviation (0.1*sd) for better visualization. (b) SPCC and MSE scores across different methods for the IDC samples. We report the average scores and scaled standard deviation (0.1*sd) for better visualization. (c) UMAP visualization from the testing set of brain to compare the generated results between Patho-R1 (base) and \method{} colored by data sources. (d) UMAP visualization from the testing set of IDC to compare the generated results between Patho-R1 (base) and \method{} colored by data sources. (e) Comparison of expression profiles between generated data and real data based on the brain tissue. (f) Comparison of expression profiles between generated data and real data based on the IDC samples.}
    \label{fig:expr bench}
\end{figure}

Figures \ref{fig:expr bench} (a) and (b) demonstrate that \method{} outperforms all baseline methods when evaluated by both SPCC and MSE across datasets from different sources, underscoring its ability to generate spot-level gene expression profiles that closely resemble measured results. Extended Data Figures \ref{supfig:gpcc example} (a) and (b) further confirm \method{}’s better performance in GPCC, highlighting its capacity to preserve gene-level heterogeneity across spatial spots. To examine the impact of base model selection on cross-modality generation, we finetuned Qwen2.5VL-7B for the same task and compared it with \method{}. As shown in Extended Data Figures \ref{supfig:gpcc example} (c) and (d), \method{}, which was built on a pathology-knowledge-enhanced VLM, outperformed the finetuned Qwen-series model. We also emphasize the importance of task-specific finetuning, supported by the clear performance gap between the unadapted base model and \method{} in generating high-quality expression profiles. UMAP visualizations of the generated profiles (Figures \ref{fig:expr bench} (c) and (d)) show that outputs from \method{} are more structured and closely aligned with reference profiles compared to those from the base model. This observation is further validated by cluster-level heatmaps of gene expression patterns in brain (Figure \ref{fig:expr bench} (e)) and IDC (Figure \ref{fig:expr bench} (f)) datasets, where \method{} more accurately recapitulates the biological signal present in the ground truth data. Collectively, these findings demonstrate that the effectiveness of \method{} in cross-modality generation arises from both the choice of a pathology-informed base model and targeted task-specific finetuning. With these advantages, \method{} represents a promising approach for generating in-silico or unseen expression profiles directly from histopathology images, thereby providing molecular-level insights into disease phenotypes.

%% file: section_folder/discussion.tex
Advances in artificial intelligence technology have endowed computational pathology with new capabilities, while the application-level focus on decision-making processes also places higher demands on the capabilities of computational pathology models. Moreover, current research lacks validation and investigation into how AI models collaborate with experts and pathologists, and the modalities of integrated information of these models remain confined to text and images. Therefore, designing an efficient AI assistant for pathology research and diagnosis holds significant practical importance. 

\textcolor{red}{Here we present \method{}, an AI copilot designed to revolutionize computational pathology and accelerate clinical disease diagnosis through a unified, multi-task framework with an automatic router for intelligent solution selection. By harnessing reinforcement learning to finetune a reasoning-enhanced VLM, \method{} achieves robust generalization and interpretable reasoning for pathology visual question answering, which shows the capabilities with direct implications for supporting pathologists in real-world diagnostic workflows. A rigorous training strategy selection and a self-verification/correction pipeline further ensure that the system produces reliable, high-quality outputs even when correcting imperfect annotations from domain experts, raising the bar for trustworthy AI-assisted diagnosis. Beyond question answering, \method{} integrates a summarization-enhanced VLM for image captioning and holistic tissue understanding, enabling richer, human-readable insights from complex histopathological images that can inform clinical reporting and decision-making. Moreover, \method{} pioneers the direct generation of spatial transcriptomic data from routine histology regions of interest, bridging the gap between morphology and molecular profiling, and opening a new frontier for multimodal data integration in precision medicine. Together, these capabilities position \method{} not only as a powerful research platform, but as a clinically meaningful tool with the potential to enhance diagnostic accuracy, reduce pathologist burden, and improve patient outcomes.}

Our experimental results show that \method{} works as a state-of-the-art method in several tasks by comparing it with advanced VLMs from general domains, medical domains, and pathology domains. \method{} can also produce more reliable reasoning paths for disease diagnosis and feature analysis. \method{} also successfully identifies the incorrect information existing in pathologists' answers and reasoning processes and provides the correction suggestions as well as corrected answers within a reasonable response time. Finally, \method{} works as a strong generator for image caption and transcriptomic information, which supports its capacity in understanding ROI-level information and integrating bimolecular information with a multi-task system. 

There are also limitations of the current implementation of \method{}. First, the improvement of base VLMs will affect the choices of components in this system, and thus, we expect to see regular model updates. Second, our task selection process relies on a trained LLM as a router, which might be substituted with a mixture-of-expert setting. \textcolor{red}{We also found that in rare cases, the model's reasoning process and its conclusion may not align (Extended Data Figure \ref{supfig:error analysis} (a)). Sometimes, the model's reasoning process is wrong and then leads to the wrong conclusions (Extended Data Figure \ref{supfig:error analysis} (b)).} However, these errors could be identified by physicians. Finally, we have not considered the privacy issues involved in pathological image analysis. Although we have made every effort to ensure that personal privacy information is not used for training, exploring defenses against attacks in this area is also important. \textcolor{red}{In the future, we will work on these directions to bring the system alive and improve its capacity and robustness for clinical usage, generating spatial transcriptomics with large resolution (e.g. ROI to WSI), integrating multi-omic data including epigenomic information with spatial resolution, as well as other enhanced approaches to build a stronger virtual assistant for diagnosis and analysis.}

%% file: section_folder/methods.tex
\textbf{Problem definition.} In this manuscript, we aim to construct a pathology-expert-level visual language model $\mathcal{M}()$ which accepts text prompts $T$ and pathology image $P$ as inputs. The outputs of our model follow the instructions and information provided in $T$ and $P$. To train $\mathcal{M}()$, we collect a dataset $D_{p}=\{(T_1,P_1),...,(T_n, P_n)\}_1^n$ with $n$ items for training, and transfer the trained model to various downstream applications. 

\textbf{Constructing \method{} as a system.} To enhance our system's multitasking capabilities, we adopted a method commonly used in current basic model development, namely training a language model-based router ($\mathcal{R}()$) according to tasks and requirements. This router accepts questions as input data and outputs the model it selects to solve specific problems. The advantage of this design is to unify the \method{} as a system for various downstream applications in digital pathology, and select the solution that best meets needs to save costs and improve model capabilities. We mark the best solution settings (one of the following choices: Reinforcement Learning (RL) \cite{sheng2024hybridflow}, Supervised FineTuning (SFT) \cite{zheng-etal-2024-llamafactory}, and test-time verification and correction (TTVC) inspired by test-time scaling (TTS) \cite{snell2024scaling}) of each question, and train $\mathcal{R}$ with questions and choices. Here, RL is used for solving questions that require reasoning, and SFT is used for summarization and cross-modality generation. Since AI Copilot needs interactions with physicians, TTVC is used for tasks requiring human-AI collaboration. Current base model of \method{} is Patho-R1-7B \cite{zhang2025patho}, which is selected after carefully comparing it with different LMMs such as Qwen2.5VL-7B \cite{team2024qwen2}, Qwen2.5VL-3B \cite{team2024qwen2}, MedVLThinker-7B \cite{huang2025medvlthinker}, PathGen-LLaVA-13B \cite{sunpathgen},  InternVL3-8B \cite{zhu2025internvl3}, and MedGemma-4B \cite{sellergren2025medgemma}.

\textbf{Empowering \method{} with reasoning capacities.} To enhance the reasoning capabilities of \method{} for complex pathological analysis, we adopt Group Relative Policy Optimization (GRPO), an efficient reinforcement learning algorithm that forgoes the critic model used in traditional PPO \cite{shao2024deepseekmath}. 

For each pathological query $q$, GRPO samples a group of $G$ outputs $\{o_1, o_2, \ldots, o_G\}$ from the current policy $\pi_\theta$ and optimizes the following objective:

\begin{equation}
J_{GRPO}(\theta) = \mathbb{E}\left[q \sim P(Q), \{o_i\}_{i=1}^G \sim \pi_{\theta_{old}}(O|q)\right] \left[\frac{1}{G}\sum_{i=1}^G \frac{1}{|o_i|}\sum_{t=1}^{|o_i|} \hat{A}_{i,t} \frac{\pi_\theta(o_{i,t}|q,o_{i,<t})}{\pi_{\theta_{old}}(o_{i,t}|q,o_{i,<t})} - \beta D_{KL}(\pi_\theta||\pi_{ref})\right],
\end{equation}

where the key novelty lies in the group-relative advantage estimation:
\begin{equation}
\hat{A}_{i,t} = \frac{r_i - \text{mean}(\mathbf{r})}{\text{std}(\mathbf{r})}.
\end{equation}

Here, $\mathbf{r} = \{r_1, r_2, \ldots, r_G\}$ represents the reward scores for all outputs in the group, obtained from a reward model trained on the quality of pathological reasoning. This group-relative formulation eliminates the need for a separate value function $V_\psi$ required in PPO, significantly reducing computational overhead while maintaining training stability.

For GRPO, the reward of question $i$ is:

\begin{equation}
r_i = r(\hat{y}_i, y_i)= \begin{cases}1, & \text { is\_equivalent }(\hat{y}_i, y_i) \\ 0, & \text { otherwise }\end{cases},
\end{equation}
where $\hat{y}$ and $y$ represent the model outputs and observed answers, respectively. is\_equivalent() is a function used to determine if the answer is correct or not.

In our ablation studies, we also consider introducing open-ended questions to model training; in that case, we utilize the BLEU score as a reward. The reward for closed-ended samples is the same, but for open-ended sample j, the reward is:

\begin{equation}
r_j = r(\hat{y}_j, y_j)= \text{BLEU}(\hat{y}_j, y_j).
\end{equation}

The comparative nature of this approach aligns naturally with pathological diagnosis workflows, where medical experts simultaneously evaluate multiple diagnostic hypotheses. By learning from the relative quality of responses within each group, \method{} develops more nuanced reasoning capabilities for tasks requiring differential diagnosis, evidence synthesis, step-by-step pathological analysis, and related questions in the pathology field.

In our ablation studies, we also consider Dynamic sAmpling Policy Optimization (DAPO) as an alternative reinforcement learning algorithm. DAPO removes the KL divergence and adjusts the group-level normalization method. That is:

\begin{equation}
\begin{aligned}
\mathcal{J}_{\mathrm{DAPO}}(\theta)= & \mathbb{E}_{(q, y) \sim \mathcal{D},\left\{o_i\right\}_{i=1}^G \sim \pi_{\theta_{\text {old }}}(\cdot \mid q)} \\
& {\left[\frac{1}{\sum_{i=1}^G\left|o_i\right|} \sum_{i=1}^G \sum_{t=1}^{\left|o_i\right|} \min \left(ra_{i, t}(\theta) \hat{A}_{i, t}, \operatorname{clip}\left(ra_{i, t}(\theta), 1-\varepsilon_{\text {low }}, 1+\varepsilon_{\text {high }}\right) \hat{A}_{i, t}\right)\right] } \\
\text { s.t. } 0 & <\mid\left\{o_i \mid \text { is\_equivalent }\left(y, o_i\right)\right\} \mid<G,
\end{aligned}
\end{equation}
where

\begin{equation}
ra_{i, t}(\theta)=\frac{\pi_\theta\left(o_{i, t} \mid q, o_{i,<t}\right)}{\pi_{\theta_{\text {old }}}\left(o_{i, t} \mid q, o_{i,<t}\right)}, \quad \hat{A}_{i, t}=\frac{r_i-\operatorname{mean}\left(\left\{r_i\right\}_{i=1}^G\right)}{\operatorname{std}\left(\left\{r_i\right\}_{i=1}^G\right)} .
\end{equation}

Here $\epsilon$ is the cut-off value to avoid gradient exploding.

\textbf{Ablation studies of training framework.} To demonstrate the efficiency and optimization of our training framework for this system, we have considered several training strategies, including 1. Supervised finetuning (SFT), which collects paired data with images and queries as inputs and answers as outputs; 2. Reinforcement Learning (RL), which utilizes the same input and output data, but we train \method{} with GRPO \cite{shao2024deepseekmath} or DAPO \cite{yu2025dapo} to tackle the reasoning capacity of selected base models; 3. SFT$+$RL, which utilizes the paired data with images, queries, and reasoning paths as inputs and answers as outputs, to train \method{} with SFT and then with RL. The first step of SFT training ensures the model acquires knowledge in relevant pathological domains, while the second step of RL training enhances the model's generalization capabilities. Our base models used for ablation studies include Qwen2.5VL-7B and Patho-R1-7B.

\textbf{Using \method{} as an AI Copilot to help pathologists.} To formalize our method as an AI Assistant, we consider two case studies inspired by the Path VQA experiments with pathologists. We invite the pathologists to answer 50 questions extracted from PathMMU from the five categories, and record their answers as well as reasoning steps. Our first case is a verifier-corrector pipeline, which can detect the incorrect answers made by pathologists and generate the correct answers. Our pipeline utilizes one verifier (a VLM, default as o4-mini) to verify whether the answers and questions proposed by pathologists are correct or not. If it is justified as wrong, we will call the corrector (also a VLM, default as \method{} used for Pathology VQA) to fix it. Otherwise, the correct answer will be returned. We have a specific threshold to limit the number of epochs in this loop. Our algorithm is summarized in Algorithm 1. We define the success of a self-verification/correction system as follows: if the expert answer is correct, or the expert answer is wrong but the answer produced by this system is correct. 

The second case is a reasoning-correction pipeline. Here we have a reasoning corrector, which takes the wrong answers and reasoning paths from pathologists, and generates the correct reasoning path with the correct answer. \textcolor{red}{Since pathologists have their own availability and we intend to improve efficiency, an offline version is selected in this paper, that is, by collecting pathologists’ reasons and answers first, we then utilize \method{} to check or correct these questions.} This pipeline can detect the wrong information provided in the reasoning process and generate the correct thinking steps. These two components focus on different aspects and work together as a prototype for building an AI Copilot that can work with pathologists and be deployed in the medical system for the future study. We have provided an example of correction in the main text. Our prompts used in these two pipelines are summarized in Appendix \ref{append: prompt list}.  

{\color{red}
We also conduct human evaluations for the reasoning paths from pathologists and \method{}. Our evaluations follow five different dimensions, including Completeness, Relevance, Conciseness, Coherence, and Clarity. All scores range from 0 to 100 and a higher score means a better reasoning path. We organize a double-blinded test after collecting the scores from different pathologists and perform analysis to ensure the fairness of our results. The grading criteria are summarized below:

\begin{itemize}
    \item \textbf{Completeness}
    \begin{itemize}
        \item Measures whether the output fully addresses all required aspects of the task or question.
        \item Covers all key components requested.
        \item Does not omit critical steps, assumptions, or conclusions.
        \item Includes necessary context, explanations, or examples (when required).
    \end{itemize}

    \item \textbf{Relevance}
    \begin{itemize}
        \item Relevance evaluates how well the model output aligns with the user’s intent and stays focused on the given task.
        \item A high score reflects content that directly responds to the prompt and avoids unnecessary or unrelated information.
        \item Responses that include tangential, off-topic, or distracting material should receive lower scores, especially if such content detracts from addressing the core request.
    \end{itemize}

    \item \textbf{Conciseness}
    \begin{itemize}
        \item Conciseness assesses how efficiently the model communicates information without unnecessary verbosity.
        \item A high score indicates that the response is succinct, avoids redundancy, and includes only the level of detail appropriate for the task.
        \item Scores should be reduced for outputs that are overly long, repetitive, or padded with filler content, particularly when verbosity obscures the main message.
    \end{itemize}

    \item \textbf{Coherence}
    \begin{itemize}
        \item Coherence measures the logical organization and flow of the response.
        \item High-scoring outputs present ideas in a clear, well-structured manner, with smooth transitions and consistent reasoning throughout.
        \item Lower scores should be given when the response is disorganized, contains abrupt shifts, internal contradictions, or a progression of ideas that is difficult to follow.
    \end{itemize}

    \item \textbf{Clarity}
    \begin{itemize}
        \item Clarity evaluates how easily the response can be understood by the intended audience.
        \item A high score reflects precise language, well-formed sentences, and unambiguous explanations, with technical terms defined when necessary.
        \item Responses that are vague, confusing, or difficult to interpret should receive lower scores, especially if lack of clarity hinders comprehension of the main points.
    \end{itemize}
\end{itemize}
}

\begin{algorithm}[htbp]
\caption{Verifier-Corrector Pipeline in \method{}.}
\begin{algorithmic}[1]
\Statex \textbf{Input:} Question $Q_M$, pathology image $I_S$, human answer $O_A$, reasoning path $O_R$, verify prompt $T_V$, correct prompt $T_C$, number of iteration $N$.
\Statex \textbf{Helper Models:} Verifier $\mathcal{M}_v$ (An advanced LMM, such as O4-mini), corrector $\mathcal{M}_c$ (An pathology-specific LMM, such as \method{} with RL finetuning), concatenation function $\cdot||\cdot$.
\Statex \textbf{Output:} Corrected outputs $O_C$
\State INIT: initialize all parameters.
\If{$\mathcal{M}_v(T_V,Q_M||O_R||O_A,I_S)$ is True}
\State $O_C=O_A$
\State Return $O_C$
\EndIf
\For{$i$ in $N$ steps}
\State $O_i,R_i=\mathcal{M}_c(T_C,Q_M||O_R||O_A,I_S)$
\If{$\mathcal{M}_v(T_V,Q_M||R_i||O_i,I_S)$ is True}
    \State $O_C=O_i$
    \State Return $O_C$
\Else
    \State $O_R = R_i$
    \State $O_A = O_i$
\EndIf
\EndFor
\State $O_C=O_A$
\State Return $O_C$
\end{algorithmic}
\end{algorithm}

\textbf{Adapting \method{} for image summarization and cross-modality generation.} To summarize the concepts in histopathology images and further generate image caption, we finetune our base model based on the paired image-caption dataset with the corresponding training set, and we also prepare 3000 samples which are only used for testing, and we utilize Deepseek-R1 \cite{guo2025deepseek} to extract the disease state and tissue source of the testing samples based on their captions. Our finetuning step follows the setting in Instruction-Tuning implemented in Llama-factory. We construct 10 different prompts to ask \method{} for generating the image captions to reduce the bias of prompt information in the training process. 

To perform the cross-modality generation task, we also finetune our base model based on the paired image-transcriptomic profile dataset with the corresponding training datasets. We select sequence data that comes from different batches and resources, but the same tissue/disease, to build a testing dataset. To transfer the information in gene expression space to text space, we first rank the genes of each spot based on their expression profiles and select the top 100 genes to formulate them in natural language. We then train a linear regressor that takes the natural language information as inputs and original gene expression profiles as outputs based on the training dataset, which finally gives us a method to decode the language information back to gene expression levels. We finetune our base model with the same approach used in image summarization and also construct 10 different prompts to ask \method{} for generating gene expression profiles.

The prompts used in this section can also be found in Appendix \ref{append: prompt list}.

\textbf{Evaluations.} In this manuscript, we consider task-specific evaluation \cite{pedregosa2011scikit,virtanen2020scipy} and follow the settings from previous works with shared tasks.

For the evaluation of Path VQA and Human-AI collaboration tasks, we utilize accuracy as a metric. The generated answer should be precisely matched with the provided answer. A higher accuracy represents a better method.

For the evaluation of the image caption summarization task, we utilize several metrics that can measure the similarity between the generated text and the provided text. These metrics include BLEU, ROUGE-1, ROUGE-2, ROUGE-L, BERT score, and MEDCON \cite{papineni2002bleu, lin2004rouge, zhangbertscore, jain1radgraph, yim2023aci}, supported by a recent publication \cite{van2024adapted}. We also consider the average score across these metrics. Here are the descriptions:

\begin{itemize}
    \item BLEU: The BiLingual Evaluation Understudy (BLEU) score evaluates the quality of generated text by breaking both the generated output and the reference text into n-grams, then comparing the overlap between the two sets. The score ranges from 0 to 1 and is typically scaled to a range of 0 to 100, with higher values indicating better model performance.

    \item ROUGE: The Recall-Oriented Understudy for Gisting Evaluation (ROUGE) score assesses text quality by computing the F1 score from n-gram overlaps between the generated text and the reference text. In this framework, n-grams from the generated text are treated as predictions, while those from the reference text serve as labels. Precision, recall, and the F1 score are calculated using the counts of matching n-grams and their lengths. ROUGE-1 measures unigram overlap, ROUGE-2 measures bigram overlap, and ROUGE-L measures the longest common subsequence. The score ranges from 0 to 1 and is typically scaled to a range of 0 to 100, with higher values indicating better model performance.

    \item BERT: The Bidirectional Encoder Representations from Transformers (BERT) model is pre-trained on large-scale text corpora for language understanding and excels at producing rich text representations. The BERTScore metric leverages this capability by measuring the similarity between embeddings of the generated text and the reference text. The score ranges from 0 to 1 and is typically scaled to a range of 0 to 100, with higher values indicating better model performance.

    \item  MEDCON: MEDCON limits the recognized concepts and entities to the semantic groups defined in QuickUMLS \cite{soldaini2016quickumls}, including Anatomy, Chemicals, Drugs, Device, Disorders, Genes, Molecular Sequences, Phenomena, and Physiology. These concepts are extracted from both the generated text and the reference text, and the F1 score is calculated based on the overlap between the two sets. The score ranges from 0 to 1 and is typically scaled to a range of 0 to 100, with higher values indicating better model performance.
\end{itemize}

A higher score of these metrics represents a better method.

For the evaluation of the cross-modality generation task, we consider spot-level Pearson Correlation Coefficient (SPCC), gene-level PCC (GPCC), and Mean Squared Error (MSE) as metrics. Higher SPCC and GPCC scores represent a better method, while a lower MSE score represents a better method. These metrics are computed between generated gene expression profiles and observed gene expression profiles.

\textbf{Baselines.} Our baseline methods cover current state-of-the-art (SOTA) open-source LMMs based on the open source movement in scientific research and the powerful influence of open source models. Moreover, there are a few powerful closed-source models focusing on digital histopathology. We apply the access to PathChat \cite{lu2024multimodal} but have not received the authorization. These models include MedGemma-4B, Qwen2.5VL-3B, Qwen2.5VL-7B, MedVLThinker-7B, InternVL3-8B, PathGen-LLaVA-13B, and Patho-R1 (7B). MedGemma-4B is an open-source VLM released by Google based on finetuning Gemma with multimodal medical data. Qwen2.5VL-3B and Qwen2.5VL-7B are open-source VLMs from the Qwen team, Alibaba Cloud. They are trained with multimodal data in the general domain. InternVL3-8B is an open-source VLM released by OpenGVLab, and it is also trained with multimodal data from the general domain. For pathology-specific models, we consider PathGen-LLaVA-13B \cite{sunpathgen}, which is finetuned based on LLAVA 13B \cite{liu2023visual} with instruction data from PathGen; as well as Patho-R1, which has a pathology-specific image encoder and is finetuned based on Qwen2.5VL-3B with reasoning data.

For the cross-modality generation task, we also consider a task-specific baseline method, known as Cell2Sentence (1B) \cite{levine2024cell2sentence}. Cell2Sentence is finetuned with instructions and single-cell transcriptomic data from atlas-level datasets based on Pythia. This model can generate cells based on instructions.

\section{Code and Data Availability}

We utilize NCSA, YCRC, and TokyoU HPC platforms to perform experiments. To train \method{}, we utilize 32 NVIDIA H100 cores and 8 NVIDIA H200 cores for 24 hours. The CPU memory upper bound is 80GB. The codes can be found in \url{https://github.com/HelloWorldLTY/TeamPath}, and the license is the MIT license.

We will release all the pre-trained model weights after peer review. The information on the datasets used in this manuscript can be found in Supplementary Table 2. To access TCGA data, an authorized account is required. To protect personal privacy, we will not release experts' answers.

\section{Acknowledgments}

We thank Mr. Tong Ding for his suggestion on model training and task selection.

\section{Author Contributions}

T.L. and W.X. designed this study. T.L., W.X., and H.Q. ran all the experiments. H.W., P.H., M.D., M.K., and A.G.T. performed human evaluation. All authors involved in writing and reviewing. H.Z. supervised this study.

\section{Institutional Review Board (IRB) Approval.}

This project has received approval from Yale IRB, with project number 2000039055. 


%% file: section_folder/appendix.tex
\counterwithin{figure}{section}
\renewcommand{\figurename}{Extended Data Fig.}
\renewcommand\thefigure{\arabic{figure}}  

\counterwithin{table}{section}
\renewcommand{\tablename}{Extended Data Tab.}
\renewcommand\thetable{\arabic{table}}  

\section{Prompt list}
\label{append: prompt list}
The prompt used for the Pathology VQA is:

\paragraph{Your task: 
1. Think through the question step by step, enclose your reasoning process in <think>...</think> tags. 
2. Then provide the correct single-letter choice (A, B, C, D,...) inside <answer>...</answer> tags.
3. No extra information or text outside of these tags.\\}

The prompt used for the self-verifier is:

\paragraph{You are an expert in pathology. You are given a QUESTION and a PROPOSED SOLUTION. Your job is to:
        1. Break down each component of the proposed solution.
        2. Think step by step to verify if the proposed solution is correct given the question and the figure.
        3. Write a line of the form "The proposed solution is correct" or "The proposed solution is incorrect" at the end of your response based on your analysis.
        QUESTION: {question}.
        PROPOSED SOLUTION: {solution}.\\}
The prompt used for the self-corrector is:

\paragraph{You are also given a question and an analysis for the question. Your job is to outline your step-by-step thought process for deriving a correct solution and also write down the correct solution. Using this format: <think> Your step-by-step reasoning of the question and solution <$/$think><answer> Your final answer <$/$answer>
            Question: {question}
            Solution: {out\_verifier}.\\}
The prompt used for the reason corrector is:

\paragraph{You are given QUESTION, REASON, and SOLUTION. Your task is to correct the REASON and SOLUTION.
        QUESTION: {question}.
        SOLUTION: {solution}.
        REASON: {reason}
        The REASON is WRONG. Your solution: \\}
The prompts used for training \method{} for caption summary include:

\paragraph{Provide a concise pathological summary of the tissue shown in this histopathology image, highlighting any abnormal cellular or structural features in one paragraph.
Based on the visual characteristics in this image, summarize the likely histological diagnosis and key indicators leading to it in one paragraph.
Describe the main histopathological patterns visible in this image and summarize what they suggest about the tissue state in one paragraph.
Summarize the key morphological findings in this histopathology image, including any signs of malignancy, inflammation, or necrosis in one paragraph.
Generate a pathology report-style summary based solely on this histological section, mentioning tissue type, grade, and diagnostic clues in one paragraph.
Briefly summarize the clinical implications of the abnormalities visible in this histopathology image in one paragraph.
From this histopathology image, extract and summarize the most diagnostically relevant features in one paragraph.
Identify and summarize any histopathological hallmarks (e.g., mitotic figures, glandular formation, stromal invasion) present in the image in one paragraph.
Write a summary suitable for a pathology trainee explaining what this histopathology image represents and why in one paragraph.
Provide an expert-level summary of the pathological findings in this histopathology image, including your confidence in the assessment in one paragraph.\\}
The prompts used for training \method{} for cross-modality generation (using IDC as an example) include:

\paragraph{Generate a list of 100 genes in order of descending expression from one spot shown in the histopathology image in IDC disease. Cell sentence:,
Produce a list of 100 gene names in descending order of expression which represent the expressed genes from one spot shown in the histopathology image in IDC disease. Cell sentence:,
Create a ranked list of 100 genes in decreasing order of expression from one spot shown in the histopathology image in IDC disease. Cell sentence:,
List the top 100 expressed genes from one spot shown in the histopathology image in IDC disease. Cell sentence:,
Identify the highest expressed 100 genes in decreasing order of expression from one spot shown in the histopathology image in IDC disease. Cell sentence:,
Enumerate a list of 100 genes in descending order of expression from one spot shown in the histopathology image in IDC disease. Cell sentence:,
Compile a descending order list of 100 expressed genes from one spot shown in the histopathology image in IDC disease. Cell sentence:,
Present a sequence of 100 genes ordered by decreasing expression level from one spot shown in the histopathology image in IDC disease. Cell sentence:,
Generate an ordered list of 100 genes by decreasing expression level from one spot shown in the histopathology image in IDC disease. Cell sentence:,
Assemble a list of 100 genes from highest to lowest expression from one spot shown in the histopathology image in IDC disease. Cell sentence:\\}

\section{SFT vs RL: Comparison for training strategies.}
\label{append: sftrl}
How to train a mature reasoning model has always been a controversial topic, as there exist several different strategies and their performances and ranks might vary under different task settings or experiment settings \cite{chu2025sft,wang2025reinforcement,liu2025part}. Meanwhile, this kind of discussion has not been investigated in training a large reasoning model for histopathology analysis, and thus, we consider several different approaches to provide an empirical analysis to select and interpret the best combination, which might inspire future directions or different researchers.

\begin{figure}[H]
    \centering
    \includegraphics[width=1\linewidth]{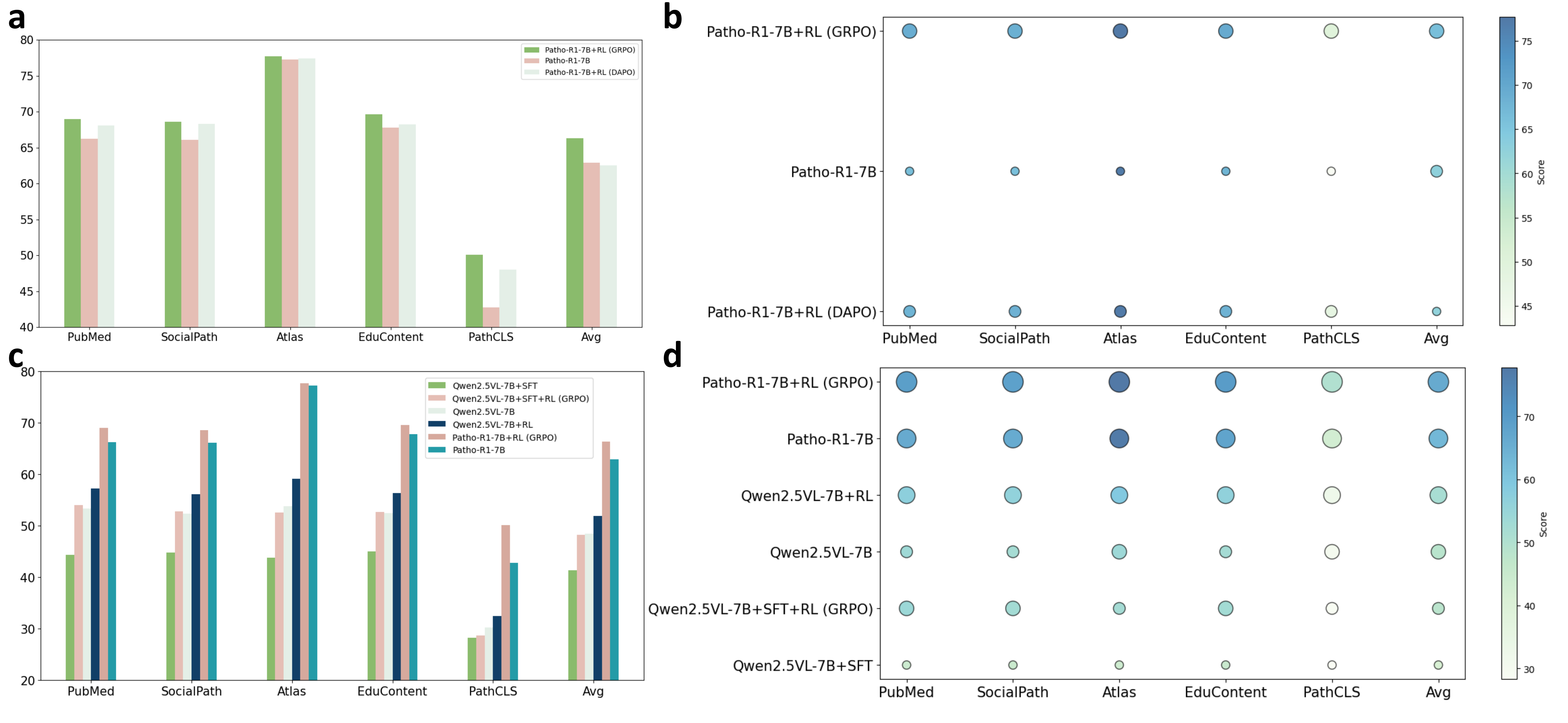}
    \caption{Training strategies optimization with different settings. The results are evaluated based on the PathMMU dataset. (a) Accuracy across different categories based on the base model and different RL strategies. (b) Accuracy and rank across different categories based on the base model and different RL strategies. (c) Accuracy across different categories based on different base models and different RL/SFT strategies. (d) Accuracy and rank across different categories based on different base models and different RL/SFT strategies.}
    \label{supfig:sftrl compare}
\end{figure}

We conduct these experiments based on different base models as well as training strategies. Our base models include Qwen2.5VL-7B, which does not contain domain-specific knowledge and Patho-R1-7B, which contains domain-specific knowledge. We also consider different training strategies, including RL (GRPO), RL (DAPO), SFT, and SFT$+$RL (GRPO). Extended Data Figures \ref{supfig:sftrl compare} (a) and (b) show that GRPO can achieve a more obvious score improvement while DAPO cannot make an improvement, which implies that a mixture of tricks does not contribute to training a multi-modal pathology reasoning model. Extended Data Figures \ref{supfig:sftrl compare} (c) and (d) show that it is important to select a base model with domain knowledge for training, and performing SFT$+$RL or direct SFT settings does not benefit this question-answer-driven task. Therefore, our optimal choice to build \method{} for pathology VQA is Patho-R1-7B+RL (GRPO).

In conclusion, our results showcase that it is important to select a good model with domain knowledge to perform training, and for LMMs that can already possess domain knowledge, directly training them using RL policies can enhance their generalization capabilities without requiring specialized SFT training (also known as cold-start training). Our important findings also align with relevant research across different fields \cite{chen2025sft,zhang2025policy}, which indirectly validates the reliability of our conclusions.

\section{Studies for training data ablation.}
\label{append: wordandchoice}
\begin{figure}[H]
    \centering
    \includegraphics[width=1\linewidth]{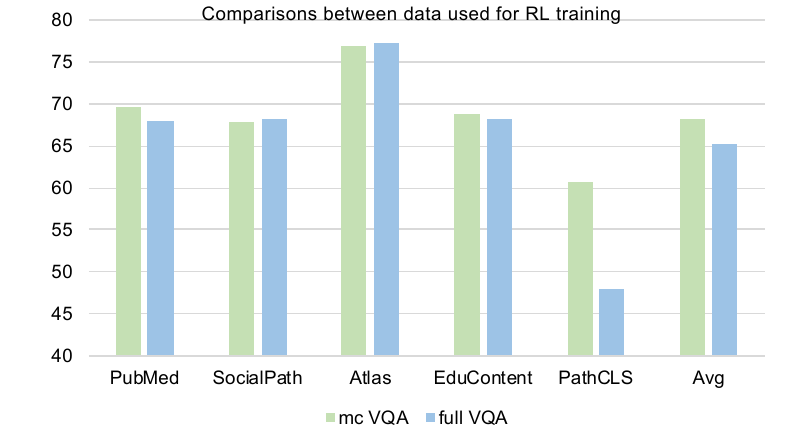}
    \caption{Comparisons for the PathMMU VQA question set with different training data. The metric is accuracy.}
    \label{supfig:openclose compare}
\end{figure}

We also investigate if incorporating more diverse data could help \method{} for generating a better reasoning path or not. To examine it, we compare the results between only using multi-choice VQA (mc VQA) data and using both multi-choice VQA and open-ended VQA (full VQA) data. Based on our evaluation results shown in Extended Data Figure \ref{supfig:openclose compare} with validation from the PathMMU dataset, using full VQA data cannot boost \method{}'s performances in generalizing the results for solving questions in the PathMMU dataset. Therefore, our optimal setting only takes mc VQA for RL training. Details of the reward design for these two different types of data are explained in the Method section.

\newpage

\section{Supplementary figures}

\begin{figure}[H]
    \centering
    \includegraphics[width=1\linewidth]{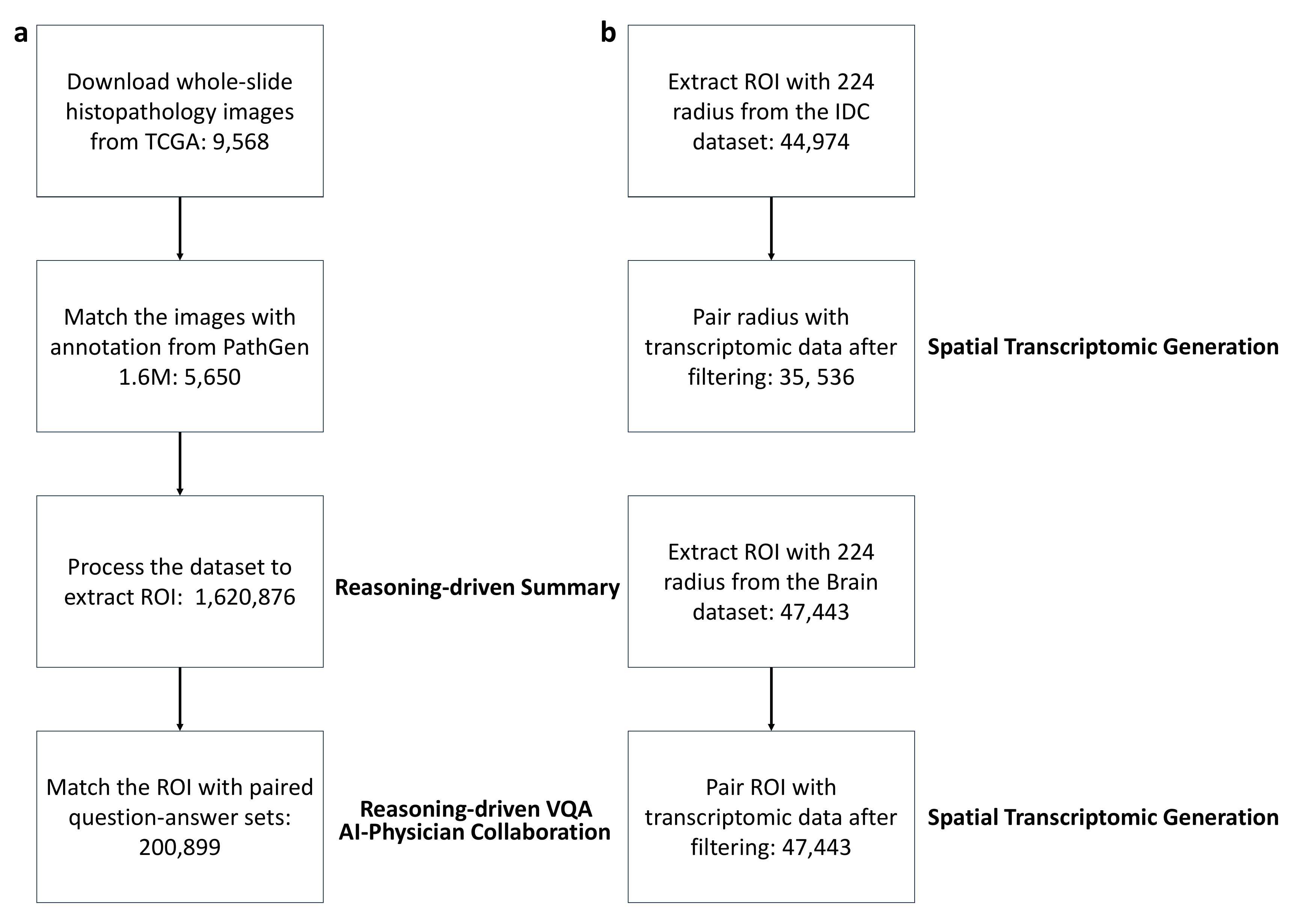}
    \caption{A flowchart of data-preprocessing used to train \method{}.}
    \label{supfig:dataflowchart}
\end{figure}

\newpage

\begin{figure}[H]
    \centering
    \includegraphics[width=1\linewidth]{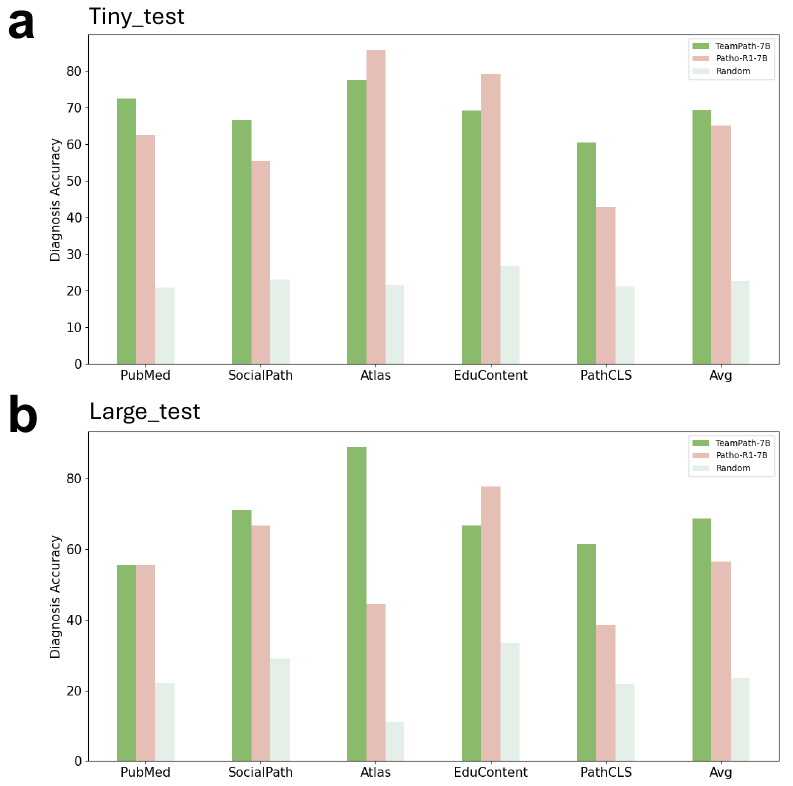}
    \caption{Accuracy of diagnosis-related questions in PathMMU. (a) Results reported based on the tiny\_test dataset. (b) Results reported based on the large\_test dataset.}
    \label{supfig:diaginfo}
\end{figure}

\newpage

\begin{figure}[H]
    \centering
    \includegraphics[width=1\linewidth]{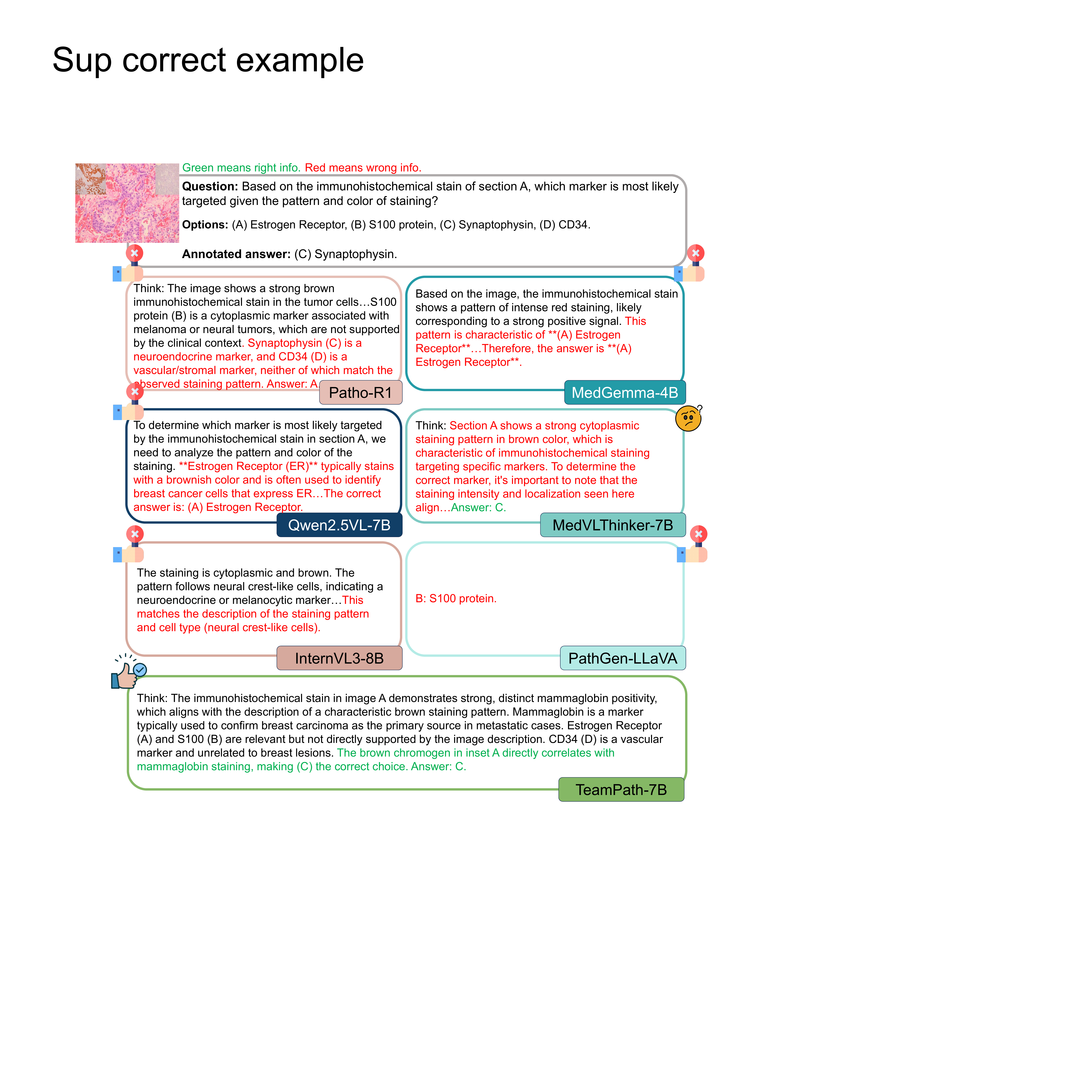}
    \caption{Case study (topic: lipoblast, which is a protein found in the presynaptic vesicles of neurons and neuroendocrine cells that plays a role in synaptic transmission) based on the outputs from different models. We highlight the correct information with green text and incorrect information with red text. For the models with errors, we consider two cases. The first case is a wrong answer, the second case is a confused reasoning path.}
    \label{supfig:casestudy2}
\end{figure}

\newpage

\newpage

\newpage

\begin{figure}[H]
    \centering
    \includegraphics[width=1\linewidth]{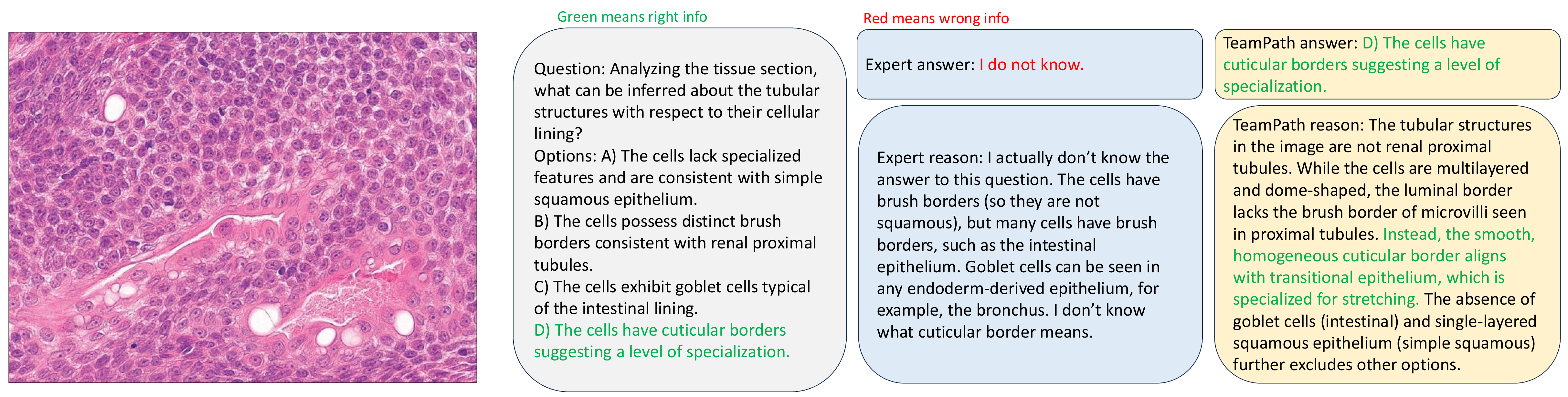}
    \caption{An example of using \method{} to correct the reasoning path from pathologists.}
    \label{supfig:correction example}
\end{figure}

\newpage

\begin{figure}[H]
    \centering
    \includegraphics[width=1\linewidth]{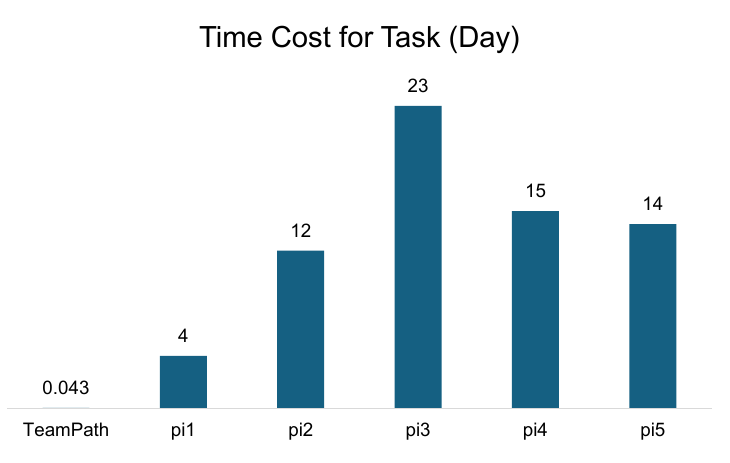}
    \caption{Comparison of the time taken by models and pathologists to solve problems. The time spent by human pathologists is calculated by subtracting the time the request was sent from the time the results were received.}
    \label{supfig:time compare}
\end{figure}

\newpage

\begin{figure}
    \centering
    \includegraphics[width=1\linewidth]{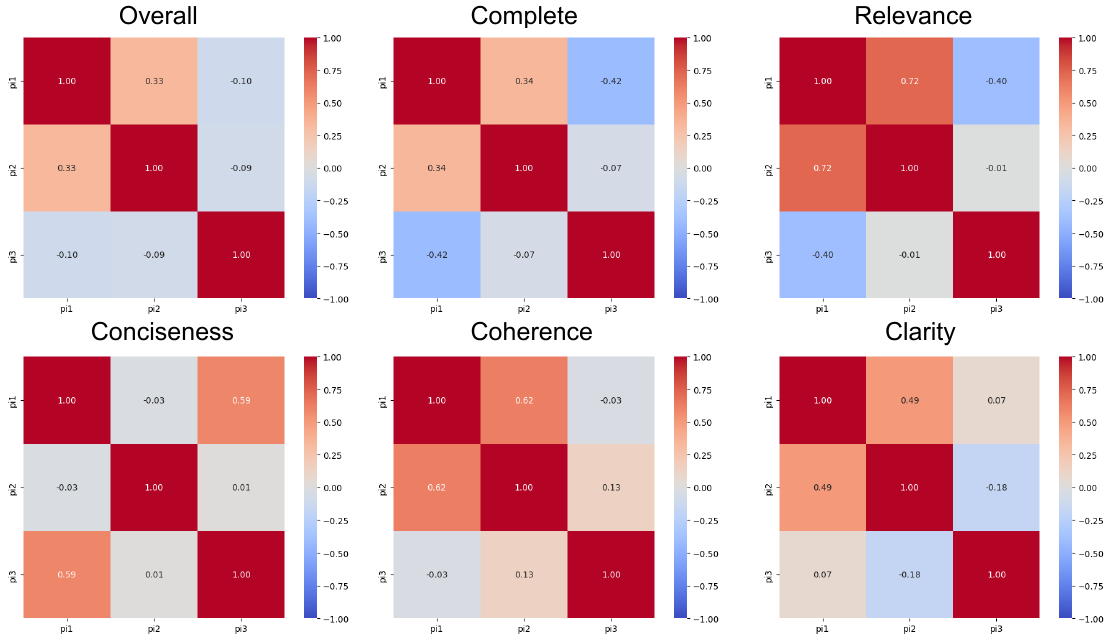}
    \caption{PCCs based on the evaluations scores from different pathologists across metrics.}
    \label{supfig:humaneval corr}
\end{figure}

\clearpage

\begin{figure}
    \centering
    \includegraphics[width=0.8\linewidth]{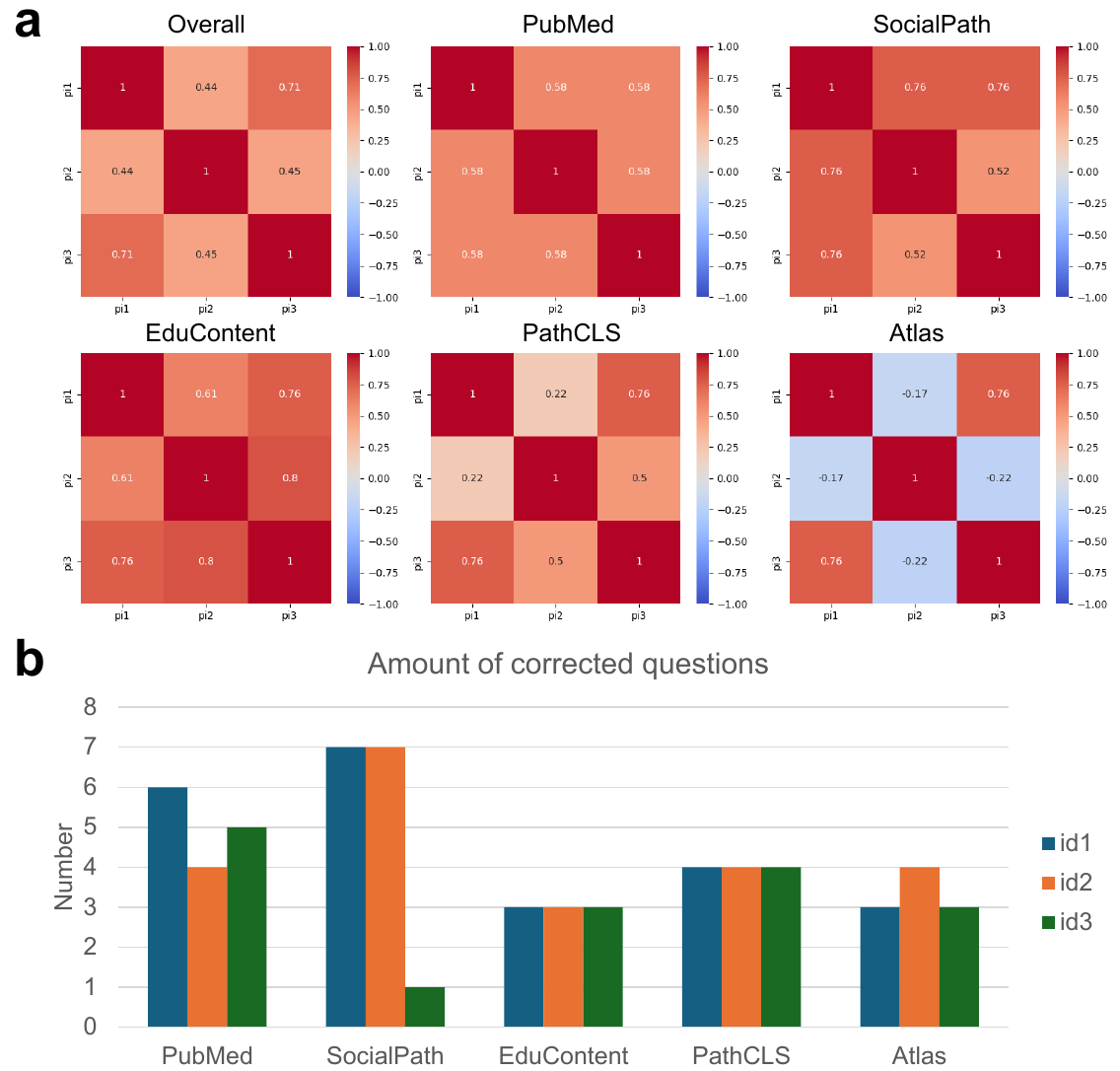}
    \caption{Results of expert feedback and \method{}-corrected information. (a) The PCCs based on the accuracy of pathologists across different question categories in PathMMU. (b) The number of corrected samples for each pathologist made by \method{}.}
    \label{supfig:human result show}
\end{figure}

\clearpage

\begin{figure}[H]
    \centering
    \includegraphics[width=1\linewidth]{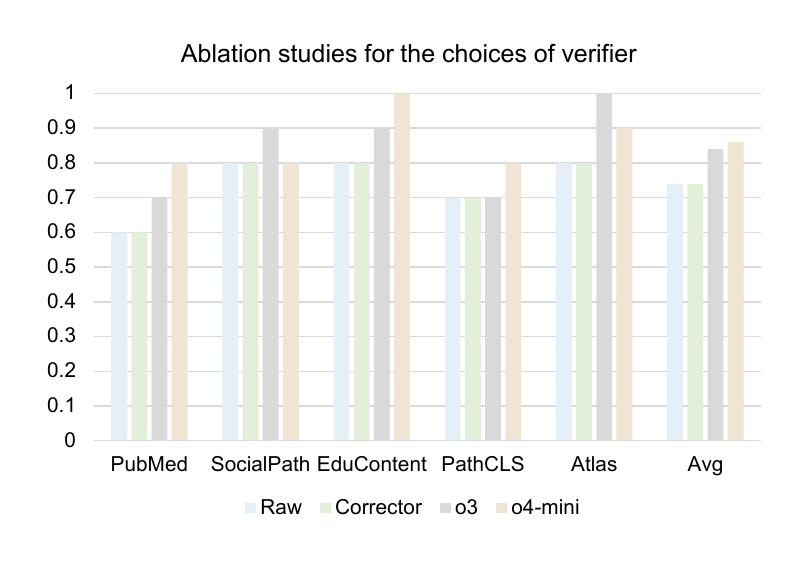}
    \caption{Ablation studies of using different models as verifiers.}
    \label{supfig:verifier abla}
\end{figure}

\clearpage

\begin{figure}
    \centering
    \includegraphics[width=1\linewidth]{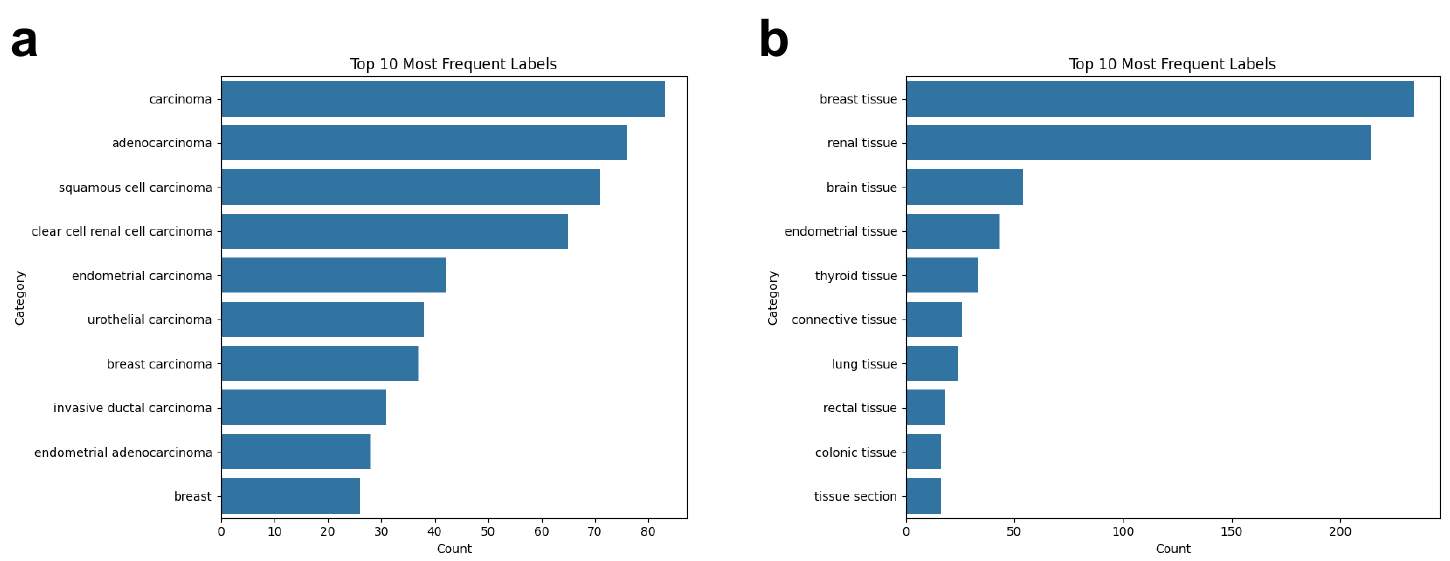}
    \caption{Distribution of image categories used for the testing of summarization performance. (a) Top 10 disease categories and their number. (b) Top 10 tissue categories and their number.}
    \label{supfig:dis and tissue category}
\end{figure}

\clearpage
\begin{figure}
    \centering
    \includegraphics[width=1\linewidth]{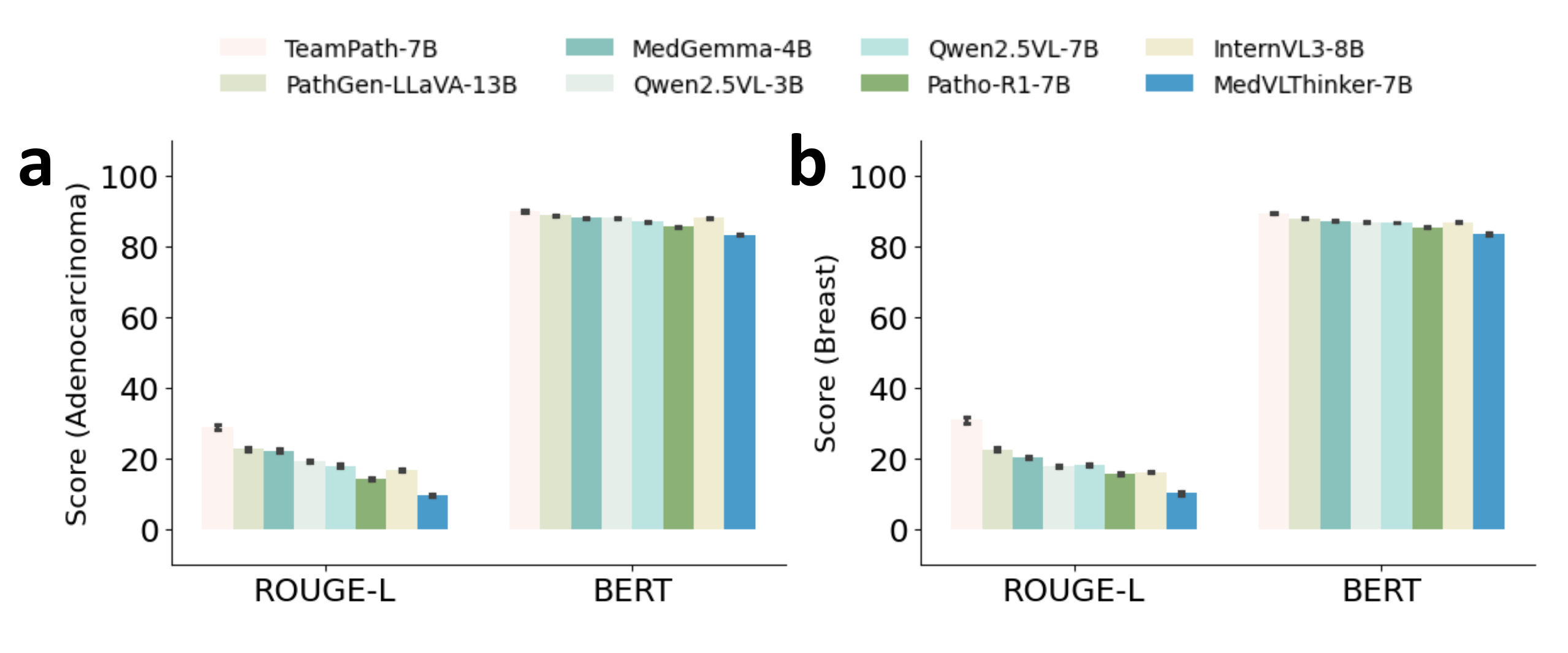}
    \caption{Summary-related evaluations by specific categories. (a) ROUGE-L and BERT scores based on samples from the selected disease across all methods. (b) ROUGE-L and BERT scores based on samples from the selected tissue across all methods.}
    \label{supfig:summary by category}
\end{figure}

\clearpage

\begin{figure}[H]
    \centering
    \includegraphics[width=1.0\linewidth]{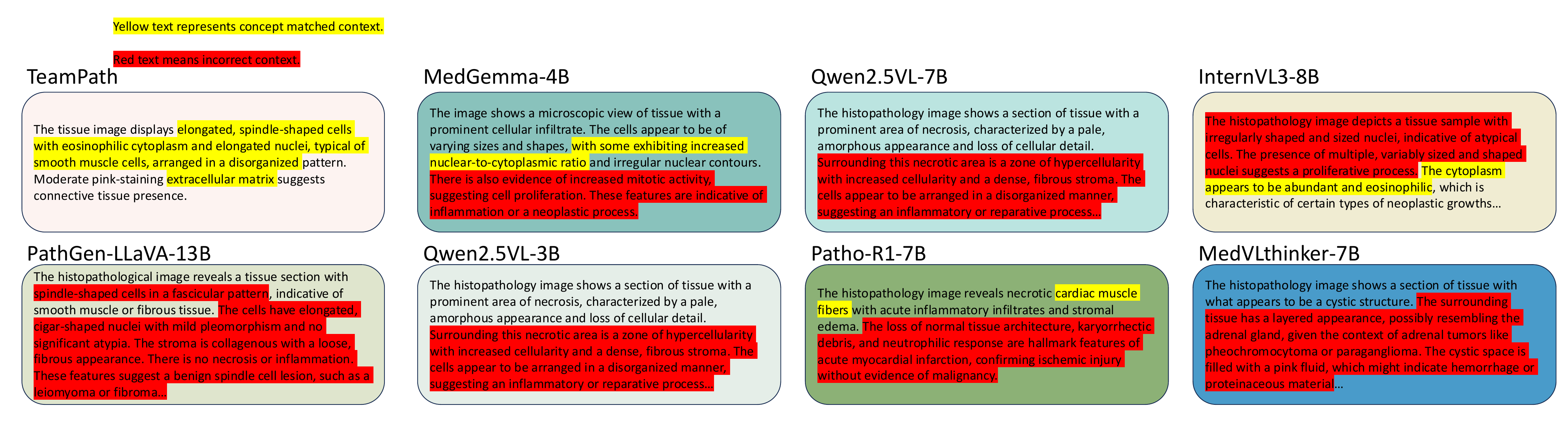}
    \caption{Examples of model outputs for caption summary tasks. We highlight both correct and incorrect information.} 
    \label{supfig:summary info}
\end{figure}

\clearpage

\begin{figure}[H]
    \centering
    \includegraphics[width=0.8\linewidth]{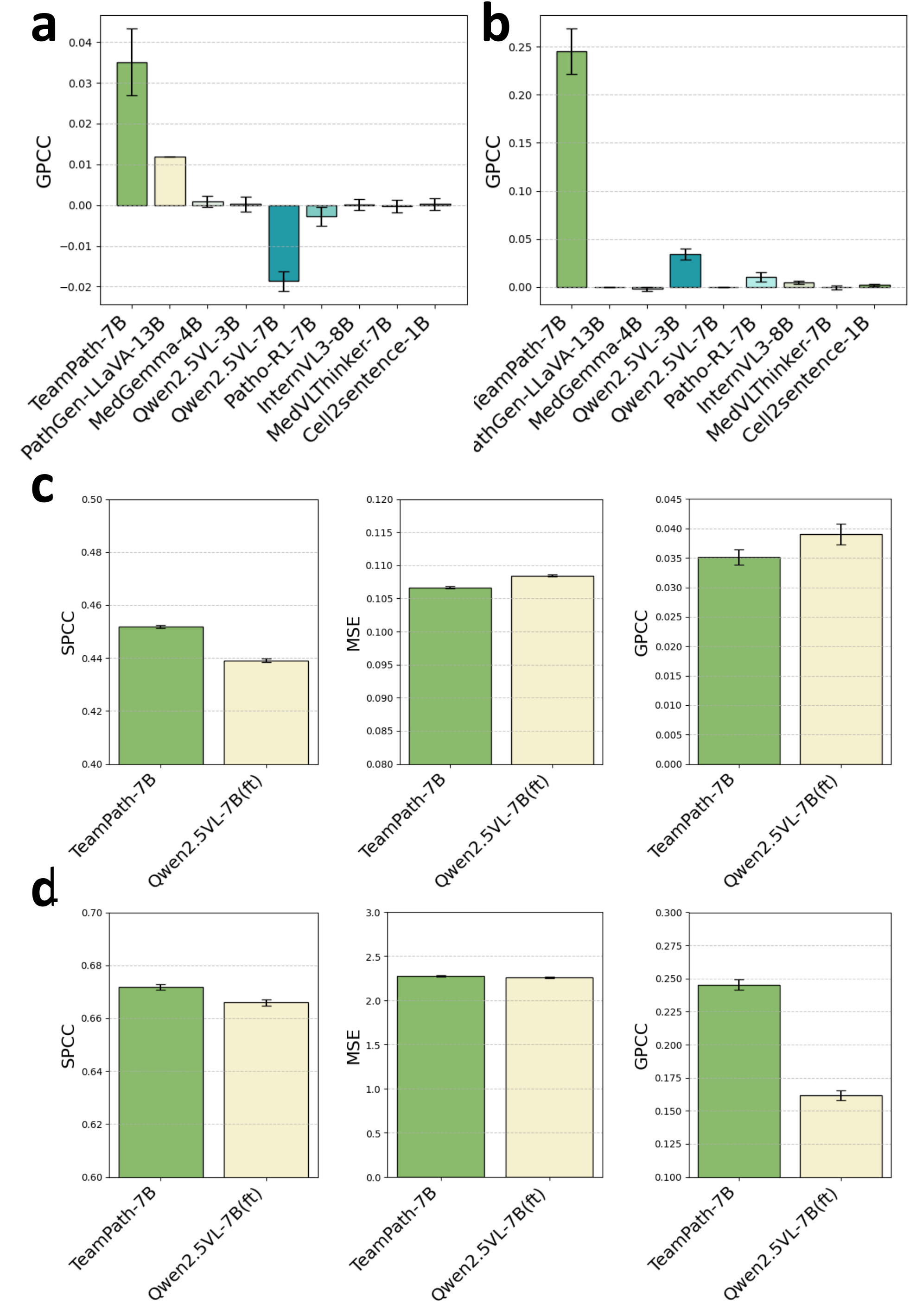}
    \caption{Extended analyses of spatial transcriptomic generation. (a) GPCC across different methods based on the brain tissue. (b) GPCC across different methods based on the IDC sample. (c) Comparison between finetuned Qwen2.5VL-7B and \method{} based on the brain tissue. (d) Comparison between finetuned Qwen2.5VL-7B and \method{} based on the IDC sample.} 
    \label{supfig:gpcc example}
\end{figure}

\newpage

\begin{figure}[H]
    \centering
    \includegraphics[width=1\linewidth]{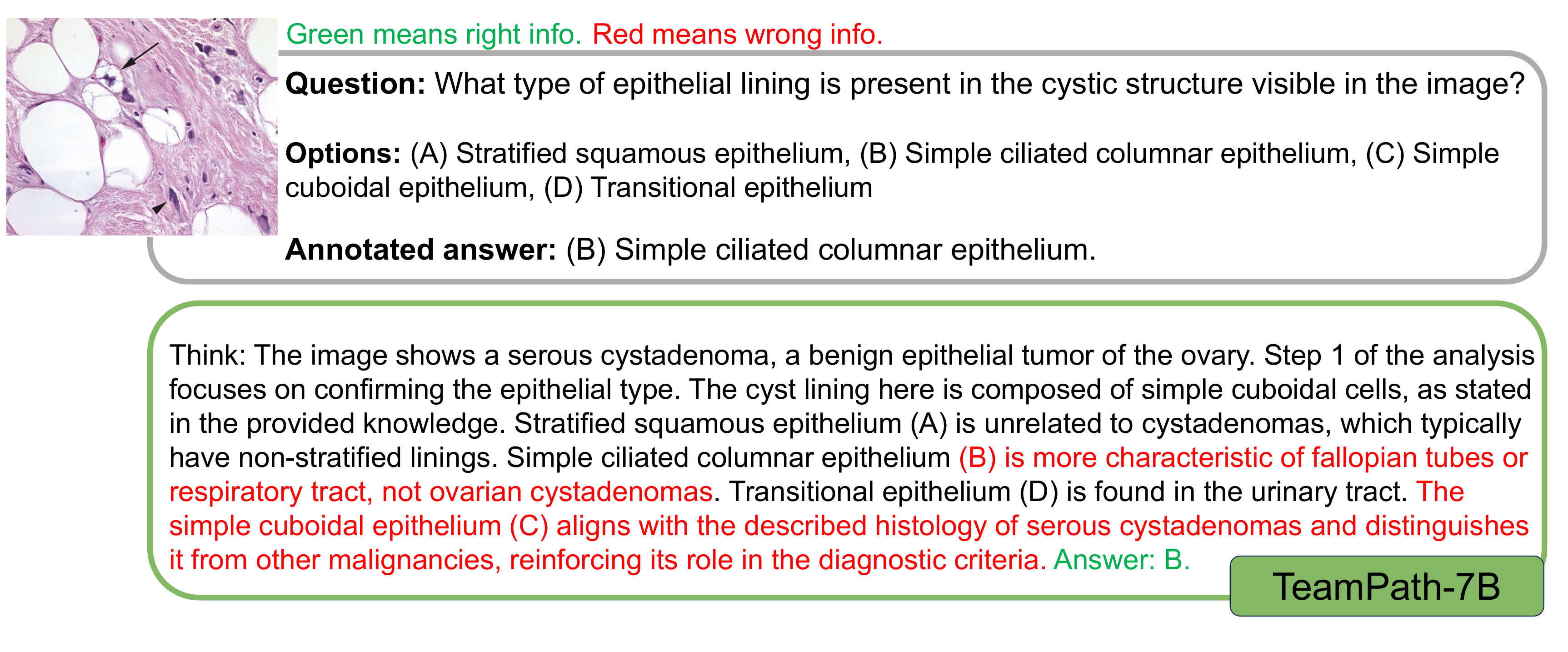}
    \caption{Case studies of \method{} output for pathology VQA with (a) an incorrect reasoning path but a correct answer and (b) an incorrect reasoning leads to an incorrect answer.}
    \label{supfig:error analysis}
\end{figure}

\newpage 

\section{Supplementary Tables}

\begin{table}[htbp]
\centering
\caption{Dataset statistics for training and testing sets.}
\begin{tabular}{lc}
\toprule
Data type & Number of samples \\
\midrule
Training\_size (only MCA) & 14,288 \\
PubMed\_test\_tiny & 281 \\
PubMed\_test & 2,787 \\
SocialPath\_test\_tiny & 216 \\
SocialPath\_test & 968 \\
Atlas\_test\_tiny & 208 \\
Atlas\_test & 799 \\
EduContent\_test\_tiny & 255 \\
EduContent\_test & 1,683 \\
PathCLS\_test & 1,632 \\
PathCLS\_test\_tiny & 177 \\
Test\_total & 9,006 \\
\bottomrule
\end{tabular}
\label{suptab: samplesize}
\end{table}

%% file: sn-bibliography.bib
@article{song2023artificial,
  title={Artificial intelligence for digital and computational pathology},
  author={Song, Andrew H and Jaume, Guillaume and Williamson, Drew FK and Lu, Ming Y and Vaidya, Anurag and Miller, Tiffany R and Mahmood, Faisal},
  journal={Nature Reviews Bioengineering},
  volume={1},
  number={12},
  pages={930--949},
  year={2023},
  publisher={Nature Publishing Group UK London}
}

@article{bera2019artificial,
  title={Artificial intelligence in digital pathology—new tools for diagnosis and precision oncology},
  author={Bera, Kaustav and Schalper, Kurt A and Rimm, David L and Velcheti, Vamsidhar and Madabhushi, Anant},
  journal={Nature reviews Clinical oncology},
  volume={16},
  number={11},
  pages={703--715},
  year={2019},
  publisher={Nature Publishing Group UK London}
}

@article{niazi2019digital,
  title={Digital pathology and artificial intelligence},
  author={Niazi, Muhammad Khalid Khan and Parwani, Anil V and Gurcan, Metin N},
  journal={The lancet oncology},
  volume={20},
  number={5},
  pages={e253--e261},
  year={2019},
  publisher={Elsevier}
}

@article{al2012digital,
  title={Digital pathology: current status and future perspectives},
  author={Al-Janabi, Shaimaa and Huisman, Andr{\'e} and Van Diest, Paul J},
  journal={Histopathology},
  volume={61},
  number={1},
  pages={1--9},
  year={2012},
  publisher={Wiley Online Library}
}

@article{zhang2024challenges,
  title={On the challenges and perspectives of foundation models for medical image analysis},
  author={Zhang, Shaoting and Metaxas, Dimitris},
  journal={Medical image analysis},
  volume={91},
  pages={102996},
  year={2024},
  publisher={Elsevier}
}

@inproceedings{
jaume2024hest,
title={{HEST}-1k: A Dataset For Spatial Transcriptomics and Histology Image Analysis},
author={Guillaume Jaume and Paul Doucet and Andrew H. Song and Ming Y. Lu and Cristina Almagro P{\'e}rez and Sophia J Wagner and Anurag Jayant Vaidya and Richard J. Chen and Drew FK Williamson and Ahrong Kim and Faisal Mahmood},
booktitle={The Thirty-eight Conference on Neural Information Processing Systems Datasets and Benchmarks Track},
year={2024},
url={https://openreview.net/forum?id=mlhFJE7PKo}
}

@misc{oai4mini,
      title={OpenAI o3 and o4-mini System Card}, 
      author={OpenAI},
      year={2025}
}

@article{pedregosa2011scikit,
  title={Scikit-learn: Machine learning in Python},
  author={Pedregosa, Fabian and Varoquaux, Ga{\"e}l and Gramfort, Alexandre and Michel, Vincent and Thirion, Bertrand and Grisel, Olivier and Blondel, Mathieu and Prettenhofer, Peter and Weiss, Ron and Dubourg, Vincent and others},
  journal={the Journal of machine Learning research},
  volume={12},
  pages={2825--2830},
  year={2011},
  publisher={JMLR. org}
}

@article{virtanen2020scipy,
  title={SciPy 1.0: fundamental algorithms for scientific computing in Python},
  author={Virtanen, Pauli and Gommers, Ralf and Oliphant, Travis E and Haberland, Matt and Reddy, Tyler and Cournapeau, David and Burovski, Evgeni and Peterson, Pearu and Weckesser, Warren and Bright, Jonathan and others},
  journal={Nature methods},
  volume={17},
  number={3},
  pages={261--272},
  year={2020},
  publisher={Nature Publishing Group}
}

@inproceedings{papineni2002bleu,
  title={Bleu: a method for automatic evaluation of machine translation},
  author={Papineni, Kishore and Roukos, Salim and Ward, Todd and Zhu, Wei-Jing},
  booktitle={Proceedings of the 40th annual meeting of the Association for Computational Linguistics},
  pages={311--318},
  year={2002}
}

@article{chen2024towards,
  title={Towards a general-purpose foundation model for computational pathology},
  author={Chen, Richard J and Ding, Tong and Lu, Ming Y and Williamson, Drew FK and Jaume, Guillaume and Song, Andrew H and Chen, Bowen and Zhang, Andrew and Shao, Daniel and Shaban, Muhammad and others},
  journal={Nature Medicine},
  volume={30},
  number={3},
  pages={850--862},
  year={2024},
  publisher={Nature Publishing Group US New York}
}

@article{lu2023towards,
  title={A visual-language foundation model for computational pathology},
  author={Lu, Ming Y and Chen, Bowen and Williamson, Drew FK and Chen, Richard J and Liang, Ivy and Ding, Tong and Jaume, Guillaume and Odintsov, Igor and Le, Long Phi and Gerber, Georg and others},
  journal={Nature medicine},
  volume={30},
  number={3},
  pages={863--874},
  year={2024},
  publisher={Nature Publishing Group US New York}
}

@article{xu2024whole,
  title={A whole-slide foundation model for digital pathology from real-world data},
  author={Xu, Hanwen and Usuyama, Naoto and Bagga, Jaspreet and Zhang, Sheng and Rao, Rajesh and Naumann, Tristan and Wong, Cliff and Gero, Zelalem and Gonz{\'a}lez, Javier and Gu, Yu and others},
  journal={Nature},
  pages={1--8},
  year={2024},
  publisher={Nature Publishing Group UK London}
}

@article{ma2024generalizablepathologyfoundationmodel,
  title={A generalizable pathology foundation model using a unified knowledge distillation pretraining framework},
  author={Ma, Jiabo and Guo, Zhengrui and Zhou, Fengtao and Wang, Yihui and Xu, Yingxue and Li, Jinbang and Yan, Fang and Cai, Yu and Zhu, Zhengjie and Jin, Cheng and others},
  journal={Nature Biomedical Engineering},
  pages={1--20},
  year={2025},
  publisher={Nature Publishing Group UK London}
}

@article{hurst2024gpt,
  title={Gpt-4o system card},
  author={Hurst, Aaron and Lerer, Adam and Goucher, Adam P and Perelman, Adam and Ramesh, Aditya and Clark, Aidan and Ostrow, AJ and Welihinda, Akila and Hayes, Alan and Radford, Alec and others},
  journal={arXiv preprint arXiv:2410.21276},
  year={2024}
}

@article{lu2024multimodal,
  title={A multimodal generative AI copilot for human pathology},
  author={Lu, Ming Y and Chen, Bowen and Williamson, Drew FK and Chen, Richard J and Zhao, Melissa and Chow, Aaron K and Ikemura, Kenji and Kim, Ahrong and Pouli, Dimitra and Patel, Ankush and others},
  journal={Nature},
  volume={634},
  number={8033},
  pages={466--473},
  year={2024},
  publisher={Nature Publishing Group UK London}
}

@article{van2024adapted,
  title={Adapted large language models can outperform medical experts in clinical text summarization},
  author={Van Veen, Dave and Van Uden, Cara and Blankemeier, Louis and Delbrouck, Jean-Benoit and Aali, Asad and Bluethgen, Christian and Pareek, Anuj and Polacin, Malgorzata and Reis, Eduardo Pontes and Seehofnerov{\'a}, Anna and others},
  journal={Nature medicine},
  volume={30},
  number={4},
  pages={1134--1142},
  year={2024},
  publisher={Nature Publishing Group US New York}
}

@misc{zhangbertscore,
  title={BERTScore: Evaluating Text Generation with BERT},
  author={Zhang, Tianyi and Kishore, Varsha and Wu, Felix and Weinberger, Kilian Q and Artzi, Yoav},
  booktitle={International Conference on Learning Representations}
}

@misc{soldaini2016quickumls,
  title={Quickumls: a fast, unsupervised approach for medical concept extraction},
  author={Soldaini, Luca and Goharian, Nazli},
  booktitle={MedIR workshop, sigir},
  pages={1--4},
  year={2016}
}

@article{xiang2025vision,
  title={A vision--language foundation model for precision oncology},
  author={Xiang, Jinxi and Wang, Xiyue and Zhang, Xiaoming and Xi, Yinghua and Eweje, Feyisope and Chen, Yijiang and Li, Yuchen and Bergstrom, Colin and Gopaulchan, Matthew and Kim, Ted and others},
  journal={Nature},
  pages={1--10},
  year={2025},
  publisher={Nature Publishing Group UK London}
}

@article{chen2025visual,
  title={A visual--omics foundation model to bridge histopathology with spatial transcriptomics},
  author={Chen, Weiqing and Zhang, Pengzhi and Tran, Tu N and Xiao, Yiwei and Li, Shengyu and Shah, Vrutant V and Cheng, Hao and Brannan, Kristopher W and Youker, Keith and Lai, Li and others},
  journal={Nature Methods},
  pages={1--15},
  year={2025},
  publisher={Nature Publishing Group}
}

@InProceedings{Chen_2025_CVPR,
    author    = {Chen, Ying and Wang, Guoan and Ji, Yuanfeng and Li, Yanjun and Ye, Jin and Li, Tianbin and Hu, Ming and Yu, Rongshan and Qiao, Yu and He, Junjun},
    title     = {SlideChat: A Large Vision-Language Assistant for Whole-Slide Pathology Image Understanding},
    booktitle = {Proceedings of the Computer Vision and Pattern Recognition Conference (CVPR)},
    month     = {June},
    year      = {2025},
    pages     = {5134-5143}
}

@article{tran2025generating,
  title={Generating dermatopathology reports from gigapixel whole slide images with HistoGPT},
  author={Tran, Manuel and Schmidle, Paul and Guo, Ruifeng Ray and Wagner, Sophia J and Koch, Valentin and Lupperger, Valerio and Novotny, Brenna and Murphree, Dennis H and Hardway, Heather D and D’Amato, Marina and others},
  journal={Nature Communications},
  volume={16},
  number={1},
  pages={1--17},
  year={2025},
  publisher={Nature Publishing Group}
}

@article{liu2025spemo,
  title={spEMO: Exploring the Capacity of Foundation Models for Analyzing Spatial Multi-Omic Data},
  author={Liu, Tianyu and Huang, Tinglin and Ding, Tong and Wu, Hao and Humphrey, Peter and Perincheri, Sudhir and Schalper, Kurt and Ying, Rex and Xu, Hua and others},
  journal={Nature Biomedical Engineering },
  pages={2025--01},
  year={2025},
  publisher={Cold Spring Harbor Laboratory}
}

@article{shaikovski2025prism2,
  title={PRISM2: Unlocking Multi-Modal General Pathology AI with Clinical Dialogue},
  author={Shaikovski, George and Vorontsov, Eugene and Casson, Adam and Viret, Julian and Zimmermann, Eric and Tenenholtz, Neil and Wang, Yi Kan and Bernhard, Jan H and Godrich, Ran A and Retamero, Juan A and others},
  journal={arXiv preprint arXiv:2506.13063},
  year={2025}
}

@article{huang2023visual,
  title={A visual--language foundation model for pathology image analysis using medical twitter},
  author={Huang, Zhi and Bianchi, Federico and Yuksekgonul, Mert and Montine, Thomas J and Zou, James},
  journal={Nature medicine},
  volume={29},
  number={9},
  pages={2307--2316},
  year={2023},
  publisher={Nature Publishing Group US New York}
}

@article{zhang2025patho,
  title={Patho-R1: A Multimodal Reinforcement Learning-Based Pathology Expert Reasoner},
  author={Zhang, Wenchuan and Zhang, Penghao and Guo, Jingru and Cheng, Tao and Chen, Jie and Zhang, Shuwan and Zhang, Zhang and Yi, Yuhao and Bu, Hong},
  journal={arXiv preprint arXiv:2505.11404},
  year={2025}
}

@article{xu2025discovering,
  title={Discovering Pathology Rationale and Token Allocation for Efficient Multimodal Pathology Reasoning},
  author={Xu, Zhe and Jin, Cheng and Wang, Yihui and Liu, Ziyi and Chen, Hao},
  journal={arXiv preprint arXiv:2505.15687},
  year={2025}
}

@article{wu2025pathvlm,
  title={PathVLM-R1: A Reinforcement Learning-Driven Reasoning Model for Pathology Visual-Language Tasks},
  author={Wu, Jianyu and Yang, Hao and Zeng, Xinhua and He, Guibing and Chen, Zhiyu and Li, Zihui and Zhang, Xiaochuan and Ma, Yangyang and Fang, Run and Liu, Yang},
  journal={arXiv preprint arXiv:2504.09258},
  year={2025}
}

@article{sheng2024hybridflow,
  title   = {HybridFlow: A Flexible and Efficient RLHF Framework},
  author  = {Guangming Sheng and Chi Zhang and Zilingfeng Ye and Xibin Wu and Wang Zhang and Ru Zhang and Yanghua Peng and Haibin Lin and Chuan Wu},
  year    = {2024},
  journal = {arXiv preprint arXiv: 2409.19256}
}

@inproceedings{zheng-etal-2024-llamafactory,
    title = "{L}lama{F}actory: Unified Efficient Fine-Tuning of 100+ Language Models",
    author = "Zheng, Yaowei  and
      Zhang, Richong  and
      Zhang, Junhao  and
      Ye, Yanhan  and
      Luo, Zheyan",
    editor = "Cao, Yixin  and
      Feng, Yang  and
      Xiong, Deyi",
    booktitle = "Proceedings of the 62nd Annual Meeting of the Association for Computational Linguistics (Volume 3: System Demonstrations)",
    month = aug,
    year = "2024",
    address = "Bangkok, Thailand",
    publisher = "Association for Computational Linguistics",
    url = "https://aclanthology.org/2024.acl-demos.38/",
    doi = "10.18653/v1/2024.acl-demos.38",
    pages = "400--410"
}

@inproceedings{
snell2024scaling,
title={Scaling {LLM} Test-Time Compute Optimally Can be More Effective than Scaling Parameters for Reasoning},
author={Charlie Victor Snell and Jaehoon Lee and Kelvin Xu and Aviral Kumar},
booktitle={The Thirteenth International Conference on Learning Representations},
year={2025},
url={https://openreview.net/forum?id=4FWAwZtd2n}
}

@article{team2024qwen2,
  title={Qwen2 technical report},
  author={Team, Qwen},
  journal={arXiv preprint arXiv:2407.10671},
  year={2024}
}

@article{zhu2025internvl3,
  title={Internvl3: Exploring advanced training and test-time recipes for open-source multimodal models},
  author={Zhu, Jinguo and Wang, Weiyun and Chen, Zhe and Liu, Zhaoyang and Ye, Shenglong and Gu, Lixin and Tian, Hao and Duan, Yuchen and Su, Weijie and Shao, Jie and others},
  journal={arXiv preprint arXiv:2504.10479},
  year={2025}
}

@article{sellergren2025medgemma,
  title={Medgemma technical report},
  author={Sellergren, Andrew and Kazemzadeh, Sahar and Jaroensri, Tiam and Kiraly, Atilla and Traverse, Madeleine and Kohlberger, Timo and Xu, Shawn and Jamil, Fayaz and Hughes, C{\'\i}an and Lau, Charles and others},
  journal={arXiv preprint arXiv:2507.05201},
  year={2025}
}

@article{guo2025deepseek,
  title={Deepseek-r1: Incentivizing reasoning capability in llms via reinforcement learning},
  author={Guo, Daya and Yang, Dejian and Zhang, Haowei and Song, Junxiao and Zhang, Ruoyu and Xu, Runxin and Zhu, Qihao and Ma, Shirong and Wang, Peiyi and Bi, Xiao and others},
  journal={arXiv preprint arXiv:2501.12948},
  year={2025}
}

@inproceedings{sunpathgen,
  title={PathGen-1.6 M: 1.6 Million Pathology Image-text Pairs Generation through Multi-agent Collaboration},
  author={Sun, Yuxuan and Zhang, Yunlong and Si, Yixuan and Zhu, Chenglu and Zhang, Kai and Shui, Zhongyi and Li, Jingxiong and Gong, Xuan and LYU, XINHENG and Lin, Tao and others},
  booktitle={The Thirteenth International Conference on Learning Representations}
}

@article{liu2023visual,
  title={Visual instruction tuning},
  author={Liu, Haotian and Li, Chunyuan and Wu, Qingyang and Lee, Yong Jae},
  journal={Advances in neural information processing systems},
  volume={36},
  pages={34892--34916},
  year={2023}
}

@inproceedings{levine2024cell2sentence,
  title={Cell2Sentence: Teaching Large Language Models the Language of Biology},
  author={Levine, Daniel and Rizvi, Syed A and L{\'e}vy, Sacha and Pallikkavaliyaveetil, Nazreen and Zhang, David and Chen, Xingyu and Ghadermarzi, Sina and Wu, Ruiming and Zheng, Zihe and Vrkic, Ivan and others},
  booktitle={International Conference on Machine Learning},
  pages={27299--27325},
  year={2024},
  organization={PMLR}
}

@article{yim2023aci,
  title={Aci-bench: a novel ambient clinical intelligence dataset for benchmarking automatic visit note generation},
  author={Yim, Wen-wai and Fu, Yujuan and Ben Abacha, Asma and Snider, Neal and Lin, Thomas and Yetisgen, Meliha},
  journal={Scientific Data},
  volume={10},
  number={1},
  pages={586},
  year={2023},
  publisher={Nature Publishing Group UK London}
}

@misc{lin2004rouge,
  title={Rouge: A package for automatic evaluation of summaries},
  author={Lin, Chin-Yew},
  booktitle={Text summarization branches out},
  pages={74--81},
  year={2004}
}

@misc{jain1radgraph,
  title={RadGraph: Extracting Clinical Entities and Relations from Radiology Reports},
  author={Jain, Saahil and Agrawal, Ashwin and Saporta, Adriel and Truong, Steven and Bui, Tan and Chambon, Pierre and Zhang, Yuhao and Lungren, Matthew P and Ng, Andrew Y and Langlotz, Curtis and others},
  booktitle={Thirty-fifth Conference on Neural Information Processing Systems Datasets and Benchmarks Track (Round 1)}
}

@inproceedings{
chu2025sft,
title={{SFT} Memorizes, {RL} Generalizes: A Comparative Study of Foundation Model Post-training},
author={Tianzhe Chu and Yuexiang Zhai and Jihan Yang and Shengbang Tong and Saining Xie and Dale Schuurmans and Quoc V Le and Sergey Levine and Yi Ma},
booktitle={Forty-second International Conference on Machine Learning},
year={2025},
url={https://openreview.net/forum?id=dYur3yabMj}
}

@article{huang2025medvlthinker,
  title={MedVLThinker: Simple Baselines for Multimodal Medical Reasoning},
  author={Huang, Xiaoke and Wu, Juncheng and Liu, Hui and Tang, Xianfeng and Zhou, Yuyin},
  journal={arXiv preprint arXiv:2508.02669},
  year={2025}
}

@article{weinstein2013cancer,
  title={The cancer genome atlas pan-cancer analysis project},
  author={Weinstein, John N and Collisson, Eric A and Mills, Gordon B and Shaw, Kenna R and Ozenberger, Brad A and Ellrott, Kyle and Shmulevich, Ilya and Sander, Chris and Stuart, Joshua M},
  journal={Nature genetics},
  volume={45},
  number={10},
  pages={1113--1120},
  year={2013},
  publisher={Nature Publishing Group}
}

@article{he2020pathvqa,
  title={Pathvqa: 30000+ questions for medical visual question answering},
  author={He, Xuehai and Zhang, Yichen and Mou, Luntian and Xing, Eric and Xie, Pengtao},
  journal={arXiv preprint arXiv:2003.10286},
  year={2020}
}

@inproceedings{sun2024pathmmu,
  title={Pathmmu: A massive multimodal expert-level benchmark for understanding and reasoning in pathology},
  author={Sun, Yuxuan and Wu, Hao and Zhu, Chenglu and Zheng, Sunyi and Chen, Qizi and Zhang, Kai and Zhang, Yunlong and Wan, Dan and Lan, Xiaoxiao and Zheng, Mengyue and others},
  booktitle={European Conference on Computer Vision},
  pages={56--73},
  year={2024},
  organization={Springer}
}

@article{chen2024stimage,
  title={Stimage-1k4m: A histopathology image-gene expression dataset for spatial transcriptomics},
  author={Chen, Jiawen and Zhou, Muqing and Wu, Wenrong and Zhang, Jinwei and Li, Yun and Li, Didong},
  journal={Advances in Neural Information Processing Systems},
  volume={37},
  pages={35796--35823},
  year={2024}
}

@article{he2020pathological,
  title={Pathological visual question answering},
  author={He, Xuehai and Cai, Zhuo and Wei, Wenlan and Zhang, Yichen and Mou, Luntian and Xing, Eric and Xie, Pengtao},
  journal={arXiv preprint arXiv:2010.12435},
  year={2020}
}

@inproceedings{mialon2023gaia,
  title={Gaia: a benchmark for general ai assistants},
  author={Mialon, Gr{\'e}goire and Fourrier, Cl{\'e}mentine and Wolf, Thomas and LeCun, Yann and Scialom, Thomas},
  booktitle={The Twelfth International Conference on Learning Representations},
  year={2023}
}

@article{liu2025towards,
  title={Towards artificial intelligence research assistant for expert-involved learning},
  author={Liu, Tianyu and Han, Simeng and Luo, Xiao and Wang, Hanchen and Lu, Pan and Zhu, Biqing and Wang, Yuge and Li, Keyi and Chen, Jiapeng and Qu, Rihao and others},
  journal={arXiv preprint arXiv:2505.04638},
  year={2025}
}

@article{song2024analysis,
  title={Analysis of 3D pathology samples using weakly supervised AI},
  author={Song, Andrew H and Williams, Mane and Williamson, Drew FK and Chow, Sarah SL and Jaume, Guillaume and Gao, Gan and Zhang, Andrew and Chen, Bowen and Baras, Alexander S and Serafin, Robert and others},
  journal={Cell},
  volume={187},
  number={10},
  pages={2502--2520},
  year={2024},
  publisher={Elsevier}
}

@misc{visiumtech,
    title={Visium technology} ,
    author={10X Genomics} 
}

@article{Rannikko2025,
  author       = {Rannikko, Antti S.},
  title        = {Artificial intelligence for prostate cancer diagnostics},
  journal      = {Nature Cancer},
  year         = {2025},
  month        = sep,
  doi          = {10.1038/s43018-025-01034-w},
}

@article{
chen2025sft,
title={{SFT} or {RL}? An Early Investigation into Training R1-Like Reasoning Large Vision-Language Models},
author={Hardy Chen and Haoqin Tu and Fali Wang and Hui Liu and Xianfeng Tang and Xinya Du and Yuyin Zhou and Cihang Xie},
journal={Transactions on Machine Learning Research},
issn={2835-8856},
year={2025},
url={https://openreview.net/forum?id=wZI5qkQeDF},
note={}
}

@inproceedings{
zhang2025policy,
title={On-Policy {RL} Meets Off-Policy Experts: Harmonizing Supervised Fine-Tuning and Reinforcement Learning via Dynamic Weighting},
author={Wenhao Zhang and Yuexiang Xie and Yuchang Sun and Yanxi Chen and Guoyin Wang and Yaliang Li and Bolin Ding and Jingren Zhou},
booktitle={The Fourteenth International Conference on Learning Representations},
year={2026},
url={https://openreview.net/forum?id=dCm9bBrk5d}
}

@article{wang2025reinforcement,
  title={Reinforcement Learning Optimization for Large-Scale Learning: An Efficient and User-Friendly Scaling Library},
  author={Wang, Weixun and Xiong, Shaopan and Chen, Gengru and Gao, Wei and Guo, Sheng and He, Yancheng and Huang, Ju and Liu, Jiaheng and Li, Zhendong and Li, Xiaoyang and others},
  journal={arXiv preprint arXiv:2506.06122},
  year={2025}
}

@article{liu2025part,
  title={Part I: Tricks or Traps? A Deep Dive into RL for LLM Reasoning},
  author={Liu, Zihe and Liu, Jiashun and He, Yancheng and Wang, Weixun and Liu, Jiaheng and Pan, Ling and Hu, Xinyu and Xiong, Shaopan and Huang, Ju and Hu, Jian and others},
  journal={arXiv preprint arXiv:2508.08221},
  year={2025}
}

@inproceedings{
yu2025dapo,
title={{DAPO}: An Open-Source {LLM} Reinforcement Learning System at Scale},
author={Qiying Yu and Zheng Zhang and Ruofei Zhu and Yufeng Yuan and Xiaochen Zuo and YuYue and Weinan Dai and Tiantian Fan and Gaohong Liu and Juncai Liu and LingJun Liu and Xin Liu and Haibin Lin and Zhiqi Lin and Bole Ma and Guangming Sheng and Yuxuan Tong and Chi Zhang and Mofan Zhang and Ru Zhang and Wang Zhang and Hang Zhu and Jinhua Zhu and Jiaze Chen and Jiangjie Chen and Chengyi Wang and Hongli Yu and Yuxuan Song and Xiangpeng Wei and Hao Zhou and Jingjing Liu and Wei-Ying Ma and Ya-Qin Zhang and Lin Yan and Yonghui Wu and Mingxuan Wang},
booktitle={The Thirty-ninth Annual Conference on Neural Information Processing Systems},
year={2025},
url={https://openreview.net/forum?id=2a36EMSSTp}
}

@article{shao2024deepseekmath,
  title={Deepseekmath: Pushing the limits of mathematical reasoning in open language models},
  author={Shao, Zhihong and Wang, Peiyi and Zhu, Qihao and Xu, Runxin and Song, Junxiao and Bi, Xiao and Zhang, Haowei and Zhang, Mingchuan and Li, YK and Wu, Yang and others},
  journal={arXiv preprint arXiv:2402.03300},
  year={2024}
}

@article{hisaoka2014lipoblast,
  title={Lipoblast: morphologic features and diagnostic value},
  author={Hisaoka, Masanori},
  journal={Journal of UOEH},
  volume={36},
  number={2},
  pages={115--121},
  year={2014},
  publisher={University of Occupational and Environmental Health, Japan}
}

@article{zhang2024vision,
  title={Vision-language models for vision tasks: A survey},
  author={Zhang, Jingyi and Huang, Jiaxing and Jin, Sheng and Lu, Shijian},
  journal={IEEE transactions on pattern analysis and machine intelligence},
  volume={46},
  number={8},
  pages={5625--5644},
  year={2024},
  publisher={IEEE}
}

@article{liu2025deepseek,
  title={Deepseek-v3. 2: Pushing the frontier of open large language models},
  author={Liu, Aixin and Mei, Aoxue and Lin, Bangcai and Xue, Bing and Wang, Bingxuan and Xu, Bingzheng and Wu, Bochao and Zhang, Bowei and Lin, Chaofan and Dong, Chen and others},
  journal={arXiv preprint arXiv:2512.02556},
  year={2025}
}
